\documentclass[12pt, peerreview , onecolumn]{IEEEtran}
\usepackage[margin=2.2cm, centering]{geometry}
\usepackage{multirow}
\usepackage{booktabs}
\usepackage{lipsum}
\usepackage{amsfonts}
\usepackage{placeins}
\usepackage{array}
\usepackage{amsmath,cases,amssymb}
\usepackage{graphicx}
\usepackage{cite}
\usepackage{subfigure}
\usepackage{epsfig}
\usepackage{url}
\usepackage{algorithm}
\usepackage{subfigure}
\usepackage{algorithmic}
\usepackage{epstopdf}
\usepackage{balance}
\usepackage{color}

\newcommand{\ds}{\displaystyle}

\newcommand{\la}{\langle}
\newcommand{\ra}{\rangle}

\newcommand{\bp}{\mathbf{p}}

\newcommand{\pkk}{p_k^{(\kappa)}}
\newcommand{\thetak}{\theta^{(\kappa)}}
\newcommand{\pikk}{\pi_k^{(\kappa)}}
\newcommand{\sthetak}{\sqrt{\thetak}}
\newcommand{\pijkk}{\pi_{j(k)}^{(\kappa)}}

\newcommand{\ta}{\tilde{a}}
\newcommand{\tb}{\tilde{b}}
\newcommand{\tc}{\tilde{c}}

\newcommand{\hk}{h^{(\kappa)}}
\newcommand{\btau}{\pmb{\tau}}
\newcommand{\clB}{\cal{B}}
\newcommand{\taukk}{\tau_k^{(\kappa)}}

\ifCLASSOPTIONpeerreview

\fi

\begin{document}

\setcounter{page}{1}

\title{UAV-Enabled Communication Using NOMA}
\author{Ali A. Nasir,  Hoang D. Tuan, Trung Q. Duong and H. Vincent Poor
\thanks{A.~A.~Nasir is with the Department of Electrical Engineering, King Fahd University of Petroleum and Minerals (KFUPM), Dhahran, Saudi Arabia (Email: anasir@kfupm.edu.sa).}
\thanks{H.~D.~Tuan is with the School of Electrical and Data Engineering, University of Technology Sydney, Broadway, NSW 2007, Australia (email: Tuan.Hoang@uts.edu.au).}
\thanks{T.~Q.~Duong is with Queen's University Belfast, Belfast BT7 1NN, UK  (email: trung.q.duong@qub.ac.uk)}%
\thanks{H.~V. Poor is with the Department of Electrical Engineering, Princeton University, Princeton, NJ 08544, USA (e-mail: poor@princeton.edu).}
}
\maketitle

\begin{abstract}

Unmanned aerial vehicles (UAVs) can be deployed as flying base stations (BSs) to leverage the strength of line-of-sight  connections and effectively support the coverage and throughput of wireless communication. This paper considers a multiuser communication system, in which a single-antenna UAV-BS
serves a large number of ground users by employing non-orthogonal multiple access (NOMA). The max-min rate optimization problem is formulated under total power, total bandwidth, UAV altitude, and  antenna beamwdith constraints. The objective of max-min rate optimization is non-convex in all optimization variables, i.e. UAV altitude, transmit antenna beamwidth, power allocation and bandwidth allocation for multiple users. A path-following algorithm is proposed to solve the formulated problem. Next, orthogonal multiple access (OMA) and dirty paper coding (DPC)-based max-min rate optimization problems are formulated and respective path-following algorithms are developed to solve them. Numerical results show that NOMA  outperforms OMA  and achieves rates  similar to those attained 
by DPC. In addition, a clear rate gain is observed by jointly optimizing all the parameters rather than optimizing a subset of parameters, which confirms the desirability of their joint optimization.

\end{abstract}

\section{Introduction}

Unmanned aerial vehicles (UAVs) can assist normal communication networks by acting as flying base stations
(UAV-BSs) and taking care of traffic demand in exceptional situations, e.g., sports events, concerts, disaster position, military situations, traffic congestion, etc. \cite{ZZL16,Jiang-12-Jun-A,Yaliniz-16-Nov-A,Meetal17, Moetal17, webdrone}. UAVs can also function as temporary hotspots or relay nodes for connections between the safe area and disaster areas \cite{ZYS11, Eretal17,Ono-16-Nov-A}.  Ground users served by the UAV-BSs can expect line-of-sight (LoS) air-to-ground communication. Thus, UAV-enabled communication can be efficient in supporting the coverage and throughput of wireless communications \cite{Meetal16,ZZL16m}.

UAV-enabled communication networks have recently gained significant interests and are actively investigated in open literature. Thanks to the flexibility of UAV deployment, the coverage area, throughput, and energy efficiency of  UAV-enabled communication can be improved by UAV placement \cite{ZZ17, CD17, ZXZ18}, beamwidth control \cite{LZZ18, Heetal18}, and power allocation \cite{ZZL16, Ghetal17, Waetal18}.

Unlike conventional cellular communication, which operates in a rich scattering
environment that supports multi-antenna array transmission for  spatial diversity,  UAV-enabled downlink communication exhibits much poorer scattering and as such a single-antenna UAV is most desired. To be served by the same UAV over the same time, multiple users must share the communication bandwidth. Usually each user is assigned an individual bandwidth channel so its achievable rate is very sensitive to the number of users sharing the same bandwidth. Naturally one may think to assign a bandwidth channel to a group users but this would be not efficient because it is conventionally known that over the same transmission bandwidth, the downlink communication is only efficient when the number of transmit antennas is not less than the number of
served users. Meanwhile, non-orthogonal multiple access (NOMA) is known to simultaneously serve multiple users in non-orthogonal resources, by separating the users in the power domain \cite{Saietal13, Dietal17}. NOMA can improve the achievable rate of far users (who receive lower received signal power) by allowing the near-by users (who receive higher received signal power) to access the information intended for the far users \cite{DAP16, Nguyen-17-Dec-A}.

There are quite a few recent studies that have considered the use of NOMA to improve the performance of UAV-enabled communication system.
In \cite{Sharma-17-Dec-P}, the authors considered a UAV-BS to communicate with two ground users using NOMA and investigated their outage probability. In \cite{Rupasinghe-18-A}, the authors considered a multi-antenna UAV-BS to generate directional beams and served multiple users to maximize their outage sum rates by using NOMA and beam scanning. In \cite{Sohail-18-A}, the authors employed a UAV system and NOMA to optimize power allocation and UAV altitude
to maximize sum-rate for two users \cite{Sohail-18-A}. However, in order to achieve the maximum rate gains from UAV-enabled communication, it is important to jointly optimize multiple relevant parameters, e.g., UAV altitude, antenna beamwidth, power allocation and bandwidth allocation. To the best  authors' knowledge, this important problem,  with a NOMA setting, is still unsolved.

In this article, we consider a multiuser communication system, in which a single-antenna UAV-BS serves a large number of ground users by employing NOMA. We jointly optimize multiple parameters, e.g., the UAV's flying altitude, transmit antenna beamwidth, and the amount of power and bandwidth allocated to multiple users. Our objective is to solve the max-min rate optimization problem under total power, total bandwidth, UAV altitude and antenna beamwidth constraints. {\color{black}The objective function is  non-convex in all optimization variables, i.e., power, bandwidth, altitude, and beamwidth. In addition, it is also challenging to handle the coverage constraint, which is dependent nonlinearly on the beamwidth and UAV altitude. We tackle these challenges by using inner convex approximations and propose a path-following algorithm to solve the problem.} We also formulate orthogonal multiple access (OMA) and dirty paper coding (DPC)-based max-min rate optimization problems and develop path-following algorithms to solve them. Numerical results show that NOMA  outperforms OMA  and achieves rates similar to those attained by DPC. In addition,
we observe a clear rate gain by jointly optimizing all the parameters rather than optimizing subset of parameters, which emphasize the need of their joint optimization.

\emph{Organization}: The paper is organized as follows. Section II presents the formulation of max-min rate optimization problems. Section III describes algorithms to solve the formulated problems. Section IV evaluates the performance of our proposed algorithms using numerical examples. Finally, Section V concludes the paper.

\section{System Model and Problem statement}\label{sec:PS}
\begin{figure*}[t]
    \centering
    \includegraphics[width=0.55 \textwidth]{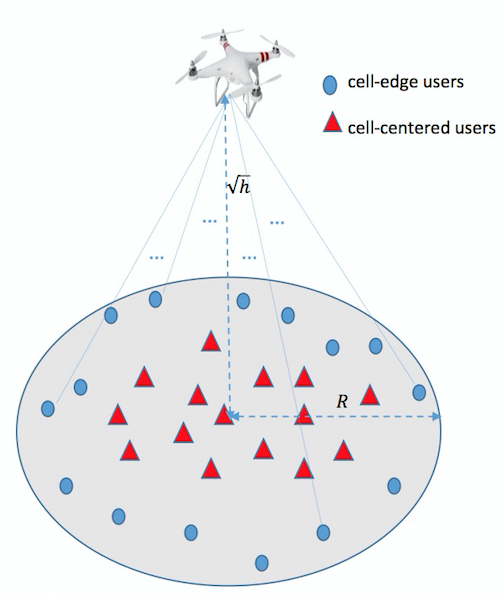}
  \caption{A system model showing UAV-BS and the ground users.}
  \label{system_model}
\end{figure*}
Let us consider that a certain out-door location 
(stadium, traffic jam, concert, etc.) is served by a single-antenna UAV
as depicted in Fig. \ref{system_model}. We assume that there are $K$ ground users in the location, such that $K/2$ users, $k\in \{1, \dots, K/2\}$, are located in closer vicinity (in terms of Euclidean distance) of the UAV, and are called ``near users" or ``cell-centered users". The remaining $K/2$ users, $k\in \{K/2+1,\dots, K\}$ are located relatively at farther distances, and are called ``far users" or ``cell-edge users". The UAV can employ NOMA to pair each near user with each of the far  users.

Let $\theta$ be the {\color{black}squared antenna beamwidth}, $h$ be the squared UAV altitude (or UAV height above ground), which must satisfy the coverage condition
\begin{equation}\label{cov}
R\leq \sqrt{h}\tan\sqrt{\theta},
\end{equation}
where $R$ is the radius of the coverage, so all users are located inside the coverage area. {\color{black}Note that we have to use $\sqrt{h}$ and $\sqrt{\theta}$ for the UAV altitude and its antenna beamwidth, respectively, as it will later on simplify the handling of non-convex coverage constraint \eqref{cov}.}
Let $g$ denote the channel power gain at a reference distance of $1$ m, $z_k = (x_k,y_k)$ denote the coordinates of user $k$ and $z_u = (x_u,y_u)$ denote the location of the UAV  projected on the horizontal ground plane. The channel power gain between the UAV and user $k$ is given by
\begin{align}\label{h}
\hbar_k(h,\theta) = \frac{g}{{\color{black}\theta}(\|z_k - z_u \|^2 + h )},
\end{align}
which assumes a free-space path loss model with path-loss exponent $2$ since users are dominated by LoS links \cite{Rupasinghe-18-A, He-18-Feb-A}.

Let $\clB$ be the total available bandwidth, which can be optimally divided among the near-by users $k\in \{1, \dots, K/2\}$, such that the bandwidth allocated for user $k$ can be written as
\begin{align}\label{wk_B}
w_k=\tau_k {\clB}, \ \ \    k\in \{1, \dots, K/2\}
\end{align}
where $0 \le \tau_k  \le 1$ is the fraction of the bandwidth allocated to the user $k$. Accordingly, each near-by user $k$ is ``assigned" a far-user $j(k) = k + K$ to share the bandwidth $w_k$.

There are a couple of transmission techniques to improve the multi-user rates. In the following, we will  formulate the multi-user rate max-min optimization problem for NOMA, DPC, and OMA.

\subsection{NOMA Problem Formulation}
To make the rate functions more appealing, we use (\ref{h}) and introduce the definitions
\[
d_k=\|z_k - z_u \|^2, k=1,\dots, K.
\]
NOMA allows user $k$ to decode the information intended for user $j(k)$ to cancel user $j(k)$'s interference
in decoding the information intended for it.  Assuming additive white Gaussian noise (AGWN) channel, the achievable rate in nats/sec/Hz
of user $k \in \{1,2,\hdots,K/2\}$, is given by
\begin{eqnarray}\label{rk}
r_k(\btau,\bp,h,\theta)&=&\ds \tau_k \ln  \left( 1 + \frac{p_k \hbar_k(h,\theta)}{\sigma_B \tau_k}    \right)\nonumber\\
  &=&\ds \tau_k \ln  \left( 1 + \frac{gp_k}{\sigma_B\tau_k\theta(d_k + h )}    \right),
k \in \{1,2,\hdots,K/2\}
\end{eqnarray}
where $\sigma_B=\sigma^2\clB$ with the noise power density
$\sigma^2$, so $\sigma_B\tau_k$ is  the noise power over the bandwidth $\tau_k\clB$, $p_k$
is the power of signal carrying the information intended for it, $ \btau \triangleq (\tau_1,\dots,\tau_{K/2})$, and $ \bp \triangleq (p_1,\dots,p_{K})$.

 The achievable rate of user $j(k)$ in nats/sec/Hz is given by
\begin{equation}\label{rjk}
r_{j(k)}(\btau,\bp,h,\theta) = \min \left\{r^{1}_{j(k)}(\btau,\bp,h,\theta),r^{2}_{j(k)}(\btau,\bp,h,\theta) \right\}
\end{equation}
where
\begin{eqnarray}\label{rk2a}
r^{2}_{j(k)}(\btau,\bp,h,\theta) &=&\ds \tau_k \ln  \left( 1 + \frac{p_{j(k)} \hbar_{j(k)}(h,\theta)}{\sigma_B\tau_k + p_k \hbar_{j(k)}(h,\theta)}    \right)\nonumber\\
&=&\tau_k \ln  \left( 1 + \frac{gp_{j(k)}}{\sigma_B\tau_k\theta(d_{j(k)}+ h )+gp_k}    \right), \ \ k \in \{1,2,\hdots,K/2\}
\end{eqnarray}
is the rate by user $j(k)$ in decoding its own message, and
\begin{eqnarray}\label{rk2b}
r^1_{j(k)}(\btau,\bp,h,\theta)& =&\ds\tau_k \ln  \left( 1 + \frac{p_{j(k)} \hbar_{k}(h,\theta)}{\sigma_B\tau_k + p_k \hbar_{k}(h,\theta)}
 \right)\nonumber\\
 &=&\tau_{k} \ln  \left( 1 + \frac{gp_{j(k)}}{\sigma_B\tau_k\theta(d_{k}+ h )+gp_k} \right), \ \
   k \in \{1,2,\hdots,K/2\}
\end{eqnarray}
is the rate by  user $k$ in decoding the user $j(k)$'s message.

The optimization problem is to find the optimal values of bandwidth allocation $\btau$, power allocation $\mathbf{p}$, UAV altitude $\sqrt{h}$, and antenna beamwidth $\sqrt{\theta}$, with the objective of maximizing the worst user's rate. It can be formulated mathematically as follows:
\begin{subequations} \label{P1}
\begin{eqnarray}
 &&\ds  \max_{\btau,\mathbf{p},h, \theta}  \ f^{\text{NOMA}}(\btau,\mathbf{p},h, \theta)\triangleq \min_{k=1,\dots, K} \  r_k^\text{NOMA}(\btau,\bp,h,\theta)
   \label{P1a}\\
   & \text{s.t.}  & (\ref{cov}),\nonumber\\
    &&h_{\min}^2\leq h\leq h_{max}^2, \theta^2_{\min}\leq\theta\leq\theta^2_{\max},\label{P1b}\\
    &&\sum_{k=1}^{K/2}\tau_k = 1,  \ \& \ \tau_k \ge 0, \ \forall k\in \{1, \dots, K/2\} \label{P1c}\\
        &&  \sum_{k=1}^{K} p_k = P, \label{P1d}
\end{eqnarray}
\end{subequations}
where
\[
r_k^\text{NOMA}(\btau,\bp,h,\theta) = \begin{cases}
r_{k}(\btau,\bp,h,\theta), & \ \  k\in\{1, \dots, K/2\}, \\
r_{j(k)}(\btau,\bp,h,\theta), & \ \  j(k)\in\{K/2+1, \dots, K\},
\end{cases}
\]
$r_{k}(\btau,\bp,h,\theta)$ is given by \eqref{rk}, $r_{j(k)}(\btau,\bp,h,\theta)$ is given by \eqref{rjk}, $P$ is the total power budget, and $\theta_{\min}$ and $\theta_{\max}$ specify the allowed range of the antenna beamwidth, i.e.,
$(0,\pi/2)$. {\color{black}It is quite challenging to solve the non-convex problem \eqref{P1} because the objective function \eqref{P1a} is  non-convex and non-linear function of four different types of variables, i.e., power, bandwidth, altitude, and beamwidth. In addition, it is also challenging to handle the coverage constraint, which  is dependent nonlinearly on the beamwidth and UAV altitude. In Section III, we will provide
an inner convex approximation-based path-following algorithm to solve this problem.}

\subsection{DPC Problem Formulation}
For two users sharing the same bandwidth, the DPC is  practical \cite{YC01,EB05,ESZ05}, under which
the rate of user $k\in\{1, \dots, K/2\}$ is defined by (\ref{rk}) while the rate of user $j(k)\in\{K/2+1, \dots, K\}$
is defined by (\ref{rk2a}). Thus, the max-min rate optimization problem under DPC  can be formulated as follows:
\begin{align}\label{PDPC}
\ds  \max_{\btau,\mathbf{p},h, \theta}  \ f^{\text{DPC}}(\btau,\mathbf{p},h, \theta)\triangleq \min_{k=1,\dots, K} \  r_k^\text{DPC}(\btau,\bp,h,\theta)\quad\mbox{s.t.}\quad
(\ref{cov}), (\ref{P1b})-(\ref{P1d}),
\end{align}
where
\[
r_k^\text{DPC}(\btau,\bp,h,\theta) = \begin{cases}
r_{k}(\btau,\bp,h,\theta), & \ \  k\in\{1, \dots, K/2\}, \\
r_{j(k)}^2(\btau,\bp,h,\theta), & \ \  j(k)\in\{K/2+1, \dots, K\},
\end{cases}
\]
$r_{k}(\btau,\bp,h,\theta)$ is given by \eqref{rk} and far-user rate $r_{j(k)}^2(\btau,\bp,h,\theta)$ is defined in (\ref{rk2a}).

\subsection{OMA Problem Formulation}
For OMA, the optimization problem can be formulated in two ways. The first way, which we term ``OMA-1" is to allocate distinct bandwidth to all users, i.e., in \eqref{wk_B}, $w_k=\tau_k {\clB}$, will be defined for $k\in \{1, \dots, K\}$. Thus, under this OMA-1, the optimization problem can be formulated as follows:
\begin{subequations}\label{P1oma1}
\begin{eqnarray}
&&  \ds  \max_{\btau,\mathbf{p},h, \theta}  \ f^{\text{OMA-1}}(\btau,\mathbf{p},h, \theta)\triangleq \min_{k=1, \dots, K}  r_k^\text{OMA-1}(\btau,\bp,h,\theta)\quad
 \text{s.t.} \quad (\ref{P1b}), (\ref{P1d}),\label{P1omaa}\\
&& \sum_{k=1}^{K}\tau_k = 1,  \ \& \ \tau_k \ge 0, \ \forall \ k\in \{1, \dots, K\} \label{P1c2}
\end{eqnarray}
\end{subequations}
where $r_k^\text{OMA-1}(\btau,\bp,h,\theta) = r_k(\btau,\bp,h,\theta)$, $\forall$ $k = \{1,\hdots,K\}$ and $r_k(\btau,\bp,h,\theta)$ is defined in \eqref{rk}.

The second option, which we term ``OMA-2", is to find optimal $K/2$ bandwidth partitions along with optimal altitude, power, and antenna beamwidth, and solve the following optimization problem:
\begin{align}\label{P1oma}
 \ds  \max_{\btau,\mathbf{p},h, \theta}  \ f^{\text{OMA-2}}(\btau,\mathbf{p},h, \theta)\triangleq \min_{k=1, \dots, K/2}  r_k^\text{OMA-2}(\btau,\bp,h,\theta)\quad
 \text{s.t.} \quad (\ref{P1b})-(\ref{P1d}),
\end{align}
where
\[
r_k^\text{OMA-2}(\btau,\bp,h,\theta) = \begin{cases}
r_{k}^\text{O}(\btau,\bp,h,\theta), & \ \  k\in\{1, \dots, K/2\}, \\
r_{j(k)}^\text{O}(\btau,\bp,h,\theta), & \ \  j(k)\in\{K/2+1, \dots, K\},
\end{cases}
\]
such that
\begin{equation}\label{rkjoma}
\begin{array}{lll}
r_k^\text{O}(\btau,\bp,h,\theta)& =&\ds
\tau_k \ln  \left( 1 + \frac{p_k \hbar_k(h,\theta)}{\sigma_B\tau_k + p_{j(k)}\hbar_k(h,\theta)} \right)\\
&=&\ds\tau_k \ln  \left( 1 + \frac{gp_k}{\sigma_B\tau_k\theta(d_k+h) + gp_{j(k)}} \right), \ \ k\in\{1,\dots, K/2\},\\
r_{j(k)}^\text{O}(\btau,\bp,h,\theta)& =&\ds
\tau_k \ln  \left( 1 + \frac{p_{j(k)} \hbar_{j(k)}(h,\theta)}{\sigma_B\tau_k + p_{k}\hbar_{j(k)}(h,\theta)} \right)\\
&=&\ds
\tau_k \ln  \left( 1 + \frac{gp_{j(k)}}{\sigma_B\tau_k\theta(d_{j(k)}+h) + gp_{k}} \right), \ \
j(k)\in\{K/2, \dots, K\}.
\end{array}
\end{equation}
\section{Algorithms}
In this section, we will solve the formulated problems in Section \ref{sec:PS}, which are non-convex optimization
problems and thus pose computational challenges.
\subsection{NOMA Algorithm}
From the definitions (\ref{rjk}), (\ref{rk2a}), and (\ref{rk2b}), one can see that the objective function (\ref{P1a}) of the
optimization problem (\ref{P1}) is a complex non-concave function. Also, constraint (\ref{cov}) is also non-convex. To
obtain a path-following computational procedure \cite{TTN16,Tametal17}, which improves a feasible point of (\ref{P1}) after each iteration and
converges to an optimal solution, we need to develop a lower-bounding concave approximation for the objective function and
also an inner convex approximation for   constraint (\ref{cov}).

Let $(\btau^{(\kappa)}, \bp^{(\kappa)}, h^{(\kappa)}, \theta^{(\kappa)})$ be a feasible point for (\ref{P1}) that is found
at the $(\kappa-1)$th iteration. With regard to the function $r_k$ in (\ref{P1}), applying inequality (\ref{ap2}) in the appendix for
\[
\tau=\tau_k, x= \sigma_B \theta /gp_k, y=\tau_k(d_k+h )
\]
and
\[
\bar{\tau}=\tau_k^{(\kappa)}, \bar{x}= \sigma_B \thetak/gp_k^{(\kappa)}, \bar{y}=\tau_k^{(\kappa)}(d_k+h^{(\kappa)})
\]
yields
\begin{eqnarray}
r_k(\btau,\mathbf{p},h,\theta)&\geq&\ds a_k^{(\kappa)}+b_k^{(\kappa)}\left(2-\frac{p_k^{(\kappa)}}{\thetak}\frac{\theta}{p_k}
-\frac{\tau_k(d_k+h )}{\tau_k^{(\kappa)}(d_k+h^{(\kappa)})}\right)-
\frac{c_k^{(\kappa)}}{\tau_k}\label{ra1}
\end{eqnarray}
where
\begin{eqnarray}
0<a_k^{(\kappa)}=2\bar{\tau}\ln(1+1/\bar{x}\bar{y}),\
0<b_k^{(\kappa)}=\frac{\bar{\tau}}{1+\bar{x}\bar{y}},\
0<c_k^{(\kappa)}=\bar{\tau}^2\ln(1+1/\bar{x}\bar{y}).\label{ra2}
\end{eqnarray}
From (\ref{ra1}) it remains to deal with
\begin{eqnarray}\label{tpk}
{\color{black}\frac{\pkk}{\thetak}\frac{\theta}{p_k}}&=&
\ds\frac{1}{4}\left( \left(\frac{\theta}{\thetak}+\frac{\pkk}{p_k} \right)^2 -
\left(\frac{\theta}{\thetak}-\frac{\pkk}{p_k} \right)^2 \right)\nonumber\\
&\leq&\ds\frac{1}{4}\left(\frac{\theta}{\thetak}+\frac{\pkk}{p_k} \right)^2\nonumber\\
&\triangleq&{\color{black}\pi_k^{(\kappa)}(\theta,p_k)}
\end{eqnarray}
and
\begin{eqnarray}
{\color{black}\ds \frac{\tau_k(d_k+h )}{\tau_k^{(\kappa)}(d_k+h^{(\kappa)})}}
&=&\ds\frac{1}{4}\left( \left(\frac{\tau_k}{\taukk}+\frac{d_k+h}{d_k+\hk}\right)^2- \left(\frac{\tau_k}{\taukk}-\frac{d_k+h}{d_k+\hk}\right)^2\right)\nonumber\\
&\leq&\ds\frac{1}{4}\left(\frac{\tau_k}{\taukk}+\frac{d_k+h}{d_k+\hk}\right)^2\nonumber\\
&\triangleq& \varphi_k^{(\kappa)}(\tau_k, h)\label{h1}
\end{eqnarray}
Therefore,
\begin{equation}\label{approx1}
r_k(\btau, \mathbf{p},h,\theta) \geq r_k^{(\kappa)}(\btau, \mathbf{p},h,\theta)
\end{equation}
for
\begin{equation}\label{brkk}
r_k^{(\kappa)}(\btau, \mathbf{p},h,\theta)\triangleq
\ds a_k^{(\kappa)}+b_k^{(\kappa)}\left(2-{\color{black}\pikk(\theta,p_k)
-\varphi_k^{(\kappa)}(\tau_k,h)}\right)-\frac{c_k^{(\kappa)}}{\tau_k},
\end{equation}
which is a concave function.
With  regard to the function $r^2_{j(k)}$, applying inequality (\ref{ap2}) in the appendix for
\[
\tau=\tau_k, x=\sigma_B \theta/gp_{j(k)}, y=\tau_k(d_{j(k)}+h )+gp_k/(\sigma_B \theta)
\]
and
\[
\bar{\tau}=\tau_k^{(\kappa)}, \bar{x}=\sigma_B \theta^{(\kappa)}/gp_{j(k)}^{(\kappa)}, \bar{y}=\tau_k^{(\kappa)}(d_{j(k)}+h^{(\kappa)})+gp_k^{(\kappa)}/(\theta^{(\kappa)} \sigma_B)
\]
yields
\begin{eqnarray}
r^2_{j(k)}(\btau, \mathbf{p},h,\theta)&\geq&\ds a_{j(k)}^{(\kappa)}+b_{j(k)}^{(\kappa)}\left(2-\frac{p_{j(k)}^{(\kappa)}}{\theta^{(\kappa)}}\frac{\theta}{p_{j(k)}}
-\frac{\tau_k(d_{j(k)}+h )+gp_k/(\sigma_B \theta)}{\tau_k^{(\kappa)}(d_{j(k)}+h^{(\kappa)})+gp_k^{(\kappa)}/(\sigma_B \thetak)}\right)\nonumber\\
&&\ds-\frac{c_{j(k)}^{(\kappa)}}{\tau_k}\label{ra5}
\end{eqnarray}
where
\begin{eqnarray}
0<a_{j(k)}^{(\kappa)}=2\bar{\tau}\ln(1+1/\bar{x}\bar{y}),\
0<b_{j(k)}^{(\kappa)}=\frac{\bar{\tau}}{1+\bar{x}\bar{y}},\
0<c_{j(k)}^{(\kappa)}=\bar{\tau}^2\ln(1+1/\bar{x}\bar{y}).\label{ra3}
\end{eqnarray}
From (\ref{ra5}), it remains to deal with
\begin{eqnarray}
{\color{black}\ds\frac{p_{j(k)}^{(\kappa)}}{\theta^{(\kappa)}}\frac{\theta}{p_{j(k)}}  }&\leq&
\ds\frac{1}{4}\left( \frac{p_{j(k)}^{(\kappa)}}{p_{j(k)}}+ \frac{\theta}{\theta^{(\kappa)}} \right)^2\nonumber\\
&\triangleq&\pi_{j(k)}^{(\kappa)}(p_{j(k)},\theta),\label{pktheta}
\end{eqnarray}
and
\begin{eqnarray}
\ds\frac{\tau_k(d_{j(k)}+h)
+gp_k/(\sigma_B \theta)}{\tau_k^{(\kappa)}(d_{j(k)}+h^{(\kappa)})+gp_k^{(\kappa)}/(\sigma_B \thetak)}&=&\nonumber\\
\ds \frac{(\tau_k/\tau_k^{(\kappa)}).[(d_{j(k)}+h)/(d_{j(k)}+h^{(\kappa)})]}
{1+gp_k^{(\kappa)}\sigma_B /\thetak\tau_k^{(\kappa)}(d_{j(k)}+h^{(\kappa)})}
+\frac{(p_k/p_k^{(\kappa)}).(\thetak/\theta) }{\sigma_B \thetak\tau_k^{(\kappa)}(d_{j(k)}+h^{(\kappa)})/gp_k^{(\kappa)}+1}&\leq&\nonumber\\
\ds \frac{1}{4}\frac{\left((\tau_k/\tau_k^{(\kappa)})+(d_{j(k)}+h)/(d_{j(k)}+h^{(\kappa)})\right)^2}
{1+gp_k^{(\kappa)}/\sigma_B \thetak\tau_k^{(\kappa)}(d_{j(k)}+h^{(\kappa)})}
+\frac{1}{4}\frac{\left((p_k/p_k^{(\kappa)})+(\thetak/\theta) \right)^2 }{\sigma_B \thetak\tau_k^{(\kappa)}(d_{j(k)}+h^{(\kappa)})/gp_k^{(\kappa)}+1}&\triangleq&\nonumber\\
\nu^{(\kappa)}_k(\tau_k,p_k, \theta).\label{pik}
\end{eqnarray}
Therefore,
\begin{equation}\label{approx2}
r^{2}_{j(k)}(\btau,\bp,h,\theta) \geq r^{2, (\kappa)}_{j(k)}(\btau,\bp,h,\theta)
\end{equation}
 for
\begin{eqnarray}\label{br2jkk}
r^{2, (\kappa)}_{j(k)}(\btau,\bp,h,\theta)&\triangleq&\ds
a_{j(k)}^{(\kappa)}+b_{j(k)}^{(\kappa)}\left(2-\pi_{j(k)}^{(\kappa)}(\theta,p_{j(k)})
-\nu^{(\kappa)}_k(\tau_k,p_k,\theta)\right)-\frac{c_{j(k)}^{(\kappa)}}{\tau_k}.
\end{eqnarray}
Analogously,
\begin{eqnarray}\label{approx3}
r^{1}_{j(k)}(\btau,\bp,h,\theta)& \geq&\ta_{j(k)}^{(\kappa)}+\tb_{j(k)}^{(\kappa)}\left(2-\frac{p_{j(k)}^{(\kappa)}}{\thetak}\frac{\theta}{p_{j(k)}}
-\frac{\tau_k(d_k+h)+gp_k/(\sigma_B \theta)}
{\tau_k^{(\kappa)}(d_k+h^{(\kappa)})+gp_k^{(\kappa)}/(\sigma_B \theta^{(\kappa)})}\right)-
\frac{\tc_{j(k)}^{(\kappa)}}{\tau_k}\nonumber\\
&\geq& r^{1, (\kappa)}_{j(k)}(\btau,\bp,h,\theta)
\end{eqnarray}
for
\begin{eqnarray}\label{br1jkk}
r^{1, (\kappa)}_{j(k)}(\btau,\bp,h,\theta)&\triangleq&\ds
\ta_{j(k)}^{(\kappa)}+\tb_{j(k)}^{(\kappa)}\left(2-\pijkk(\theta,p_{j(k)})-\tilde{\nu}^{(\kappa)}_k(\tau_k,p_k,\theta)
\right) -\frac{\tc_{j(k)}^{(\kappa)}}{\tau_k},
\end{eqnarray}
and
\begin{equation}\label{tnuk}
\tilde{\nu}^{(\kappa)}_k(\tau_k,p_k, \theta)\triangleq
\ds \frac{1}{4}\frac{\left((\tau_k/\tau_k^{(\kappa)})+(d_{k}+h)/(d_{k}+h^{(\kappa)})\right)^2}
{1+gp_k^{(\kappa)}/\sigma_B \thetak\tau_k^{(\kappa)}(d_{k}+h^{(\kappa)})}
+\frac{1}{4}\frac{\left((p_k/p_k^{(\kappa)})+(\thetak/\theta) \right)^2 }{\sigma_B \thetak\tau_k^{(\kappa)}(d_{k}+h^{(\kappa)})/gp_k^{(\kappa)}+1},
\end{equation}
and
\begin{eqnarray}
\ta_{j(k)}^{(\kappa)}=2\bar{\tau}\ln(1+1/\bar{x}\bar{y}),\
\tb_{j(k)}^{(\kappa)}=\frac{\bar{\tau}}{1+\bar{x}\bar{y}},\
\tc_{j(k)}^{(\kappa)}=\bar{\tau}^2\ln(1+1/\bar{x}\bar{y}),\label{ra4}
\end{eqnarray}
under
\[
\bar{\tau}=\tau_k^{(\kappa)}, \bar{x}=\sigma_B \thetak/gp_{j(k)}^{(\kappa)}, \bar{y}=\tau_k^{(\kappa)}(d_{k}+h^{(\kappa)})+gp_k^{(\kappa)}/(\sigma_B \thetak).
\]
A lower bounding concave function for the objective function (\ref{P1a}) is
\begin{equation}\label{lbobj}
f^{\text{NOMA}, (\kappa)}(\btau,\bp,h,\theta)=\min_{k=1,\dots,K} r_k^{\text{NOMA},(\kappa)}(\btau,\bp,h,\theta)
\end{equation}
where
\begin{equation}\label{lbnoma}
r_k^{\text{NOMA},(\kappa)}(\btau,\bp,h,\theta) = \begin{cases}
r^{(\kappa)}_{k}(\btau,\bp,h,\theta), & \ \  k\in\{1, \dots, K/2\}, \\
r^{(\kappa)}_{j(k)}(\btau,\bp,h,\theta), & \ \  j(k)\in\{K/2+1, \dots, K\},
\end{cases}
\end{equation}
and
\[
r^{(\kappa)}_{j(k)}(\btau,\bp,h,\theta) = \min \left\{r^{1,(\kappa)}_{j(k)}(\btau,\bp,h,\theta),r^{2,(\kappa)}_{j(k)}(\btau,\bp,h,\theta) \right\}.
\]
with $r^{(\kappa)}_{k}(\btau,\bp,h,\theta)$, $r^{1,(\kappa)}_{j(k)}(\btau,\bp,h,\theta)$, and $r^{2,(\kappa)}_{j(k)}(\btau,\bp,h,\theta)$  defined in \eqref{brkk}, \eqref{br1jkk}, and \eqref{br2jkk}, respectively.

It remains to deal with the non-convex constraint (\ref{cov}). From the convexity of the tangential function, 
it follows that
\begin{eqnarray}\label{ecov}
\sqrt{h}\tan\sqrt{\theta}&\geq& \sqrt{h}(\tan\sthetak+\frac{\sqrt{\theta}-\sthetak}{(\cos\sthetak)^2})\nonumber\\
&=&\ds\frac{\sin\sthetak\cos\sthetak-\sthetak}{(\cos\sthetak)^2}\sqrt{h}+\frac{\sqrt{h\theta}}{(\cos\sthetak)^2}\nonumber\\
&\geq&\ds\frac{\sin\sthetak\cos\sthetak-\sthetak}{(\cos\sthetak)^2}\left(\frac{\sqrt{h^{(\kappa)}}}{2}+
\frac{h}{2\sqrt{h^{(\kappa)}}}\right)+\frac{\sqrt{h\theta}}{(\cos\sthetak)^2}.
\end{eqnarray}
Therefore, an inner approximation of (\ref{cov}) is\footnote{$\sin\sthetak\cos\sthetak-\sthetak<0$}
\begin{equation}\label{ecovk}
R\leq \ds\frac{\sin\sthetak\cos\sthetak-\sthetak}{(\cos\sthetak)^2}\left(\frac{\sqrt{h^{(\kappa)}}}{2}+
\frac{h}{2\sqrt{h^{(\kappa)}}}\right)+\frac{\sqrt{h\theta}}{(\cos\sthetak)^2},
\end{equation}
i.e. every feasible point for the latter is also feasible for the former.

In summary, at the $\kappa$-th iteration, we solve the following convex optimization problem to generate the next
iterative feasible point $(\tau^{(\kappa+1)}, p^{(\kappa+1)}, \theta^{(\kappa+1)}, h^{(\kappa+1)})$:
\begin{eqnarray} \label{eP1k}
 \ds  \max_{\btau,\mathbf{p},h, \theta} \ f^{\text{NOMA},(\kappa)}(\btau,\bp,h,\theta)\quad
 \text{s.t.} \quad (\ref{P1b}), (\ref{P1c}), (\ref{P1d}), (\ref{ecovk}),
\end{eqnarray}

Algorithm \ref{alg1} outlines the steps to solve the max-min rate optimization problem \eqref{P1}.

\emph{\underline{Finding an initial feasible point}}: The initial feasible point $(\btau^{(0)}, \bp^{(0)}, \theta^{(0)}, h^{(0)})$ can be obtained by following the following three steps.
\begin{enumerate}
\item First, we can calculate $\btau^{(0)}$ and  $\bp^{(0)}$ by simply assuming equal power and equal bandwidth allocation, i.e., $p_k^{(0)} = P/K$, $\forall$ $k$, and $\tau_k^{(0)} = 1/(K/2)$, $\forall$ $k \in \{1,\hdots,K/2\}$.
\item We can find $\theta^{(0)}$ by fixing it to some value that satisfies $\theta^2_{\min}\leq\theta\leq\theta^2_{\max}$ in \eqref{P1b}.
\item Finally, we can find $h^{(0)}$ by solving a feasibility problem for $h$ under convex constraints $h_{\min}^2\leq h\leq h_{max}^2$ and $R\leq \sqrt{h}\tan\sqrt{\theta^{(0)}}$.
\end{enumerate}

Note that
$f^{\text{NOMA},(\kappa)}(\btau^{(\kappa+1)},\bp^{(\kappa+1)},h^{(\kappa+1)},\theta^{(\kappa+1)})>
f^{\text{NOMA},(\kappa)}(\btau^{(\kappa)},\bp^{(\kappa)},h^{(\kappa)},\theta^{(\kappa)})$ because
$(\btau^{(\kappa+1)}, \newline \bp^{(\kappa+1)},h^{(\kappa+1)},\theta^{(\kappa+1)})$ and $(\btau^{(\kappa)},  \bp^{(\kappa)},h^{(\kappa)},\theta^{(\kappa)})$ are respectively the optimal solution and a
feasible point of (\ref{eP1k}). Therefore
\begin{eqnarray}
f^{\text{NOMA}}(\btau^{(\kappa+1)},\bp^{(\kappa+1)},h^{(\kappa+1)},\theta^{(\kappa+1)})&\geq&
f^{\text{NOMA},(\kappa)}(\btau^{(\kappa+1)},\bp^{(\kappa+1)},h^{(\kappa+1)},\theta^{(\kappa+1)})\label{ine1}\\
&>&f^{\text{NOMA},(\kappa)}(\btau^{(\kappa)},\bp^{(\kappa)},h^{(\kappa)},\theta^{(\kappa)})\nonumber\\
&=&f^{\text{NOMA}}(\btau^{(\kappa)},\bp^{(\kappa)},h^{(\kappa)},\theta^{(\kappa)}),\label{ine2}
\end{eqnarray}
where (\ref{ine1}) is true because $f^{\text{NOMA},(\kappa)}$ is a lower bound of $f^{\text{NOMA}}$
while (\ref{ine2}) is true  because $f^{\text{NOMA},(\kappa)}$ matches with $f^{\text{NOMA}}$
at $(\btau^{(\kappa)},\bp^{(\kappa)},h^{(\kappa)},\theta^{(\kappa)})$, so
$(\btau^{(\kappa+1)},\bp^{(\kappa+1)},h^{(\kappa+1)},\theta^{(\kappa+1)})$ is a better feasible point than
$(\btau^{(\kappa)},\bp^{(\kappa)},h^{(\kappa)},\theta^{(\kappa)})$. As such, the sequence $\{(\btau^{(\kappa)},\bp^{(\kappa)},h^{(\kappa)},\theta^{(\kappa)})\}$ at least converges to a locally optimal solution
of (\ref{P1}) \cite{TTN16,Tametal17}.

\begin{algorithm}[t]
\begin{algorithmic}[1]
\protect\caption{NOMA-based algorithm for max-min rate optimization problem \eqref{P1}}
\label{alg1}
\global\long\def\algorithmicrequire{\textbf{Initialization:}}
\REQUIRE  Set $\kappa:=0$ and a feasible point $(\btau^{(0)}, \bp^{(0)}, \theta^{(0)}, h^{(0)})$ for constraints \eqref{cov}, \eqref{P1b}, \eqref{P1c}, and \eqref{P1d}.
\REPEAT
\STATE Solve the convex  optimization problem (\ref{eP1k}) to obtain the optimal solution $(\btau^{(\kappa+1)}, \bp^{(\kappa+1)}, \theta^{(\kappa+1)}, h^{(\kappa+1)})$.
\STATE Set $\kappa:=\kappa+1.$
\UNTIL Convergence\\
\end{algorithmic} \end{algorithm}

\subsection{DPC Algorithm}

The objective function of the DPC problem \eqref{PDPC} has  structure similar to that for the NOMA problem \eqref{P1}. The non-convex constraint \eqref{cov} can be approximated
by \eqref{ecovk}.  Thus, we can use the inequality \eqref{ap2} and  approximations \eqref{tpk}, \eqref{h1}, \eqref{pktheta}, and \eqref{pik},  to approximate the non-concave objective function in \eqref{PDPC}.
Therefore, at the $\kappa$-th iteration, we solve the following convex optimization problem to generate the next
iterative feasible point $(\tau^{(\kappa+1)}, p^{(\kappa+1)}, \theta^{(\kappa+1)}, h^{(\kappa+1)})$:
\begin{eqnarray} \label{eP1kdpc}
 \ds  \max_{\btau,\mathbf{p},h, \theta} \ f^{\text{DPC},(\kappa)}\triangleq \min_{k=1,\dots, K/2}\ r^{\text{DPC},(\kappa)}_k(\btau,\bp,h,\theta)\quad
 \text{s.t.} \quad (\ref{P1b}), (\ref{P1c}), (\ref{P1d}), (\ref{ecovk}),
\end{eqnarray}
where
\[
r_k^{\text{DPC},(\kappa)}(\btau,\bp,h,\theta) = \begin{cases}
r^{(\kappa)}_{k}(\btau,\bp,h,\theta), & \ \  k\in\{1, \dots, K/2\}, \\
r^{2,(\kappa)}_{j(k)}(\btau,\bp,h,\theta), & \ \  j(k)\in\{K/2+1, \dots, K\}.
\end{cases}
\]
Note that $r^{(\kappa)}_{k}(\btau,\bp,h,\theta)$ and $r^{2,(\kappa)}_{j(k)}(\btau,\bp,h,\theta)$ are defined in \eqref{brkk} and \eqref{br2jkk}, respectively.

Similar to Algorithm \ref{alg1}, Algorithm \ref{alg2} outlines the steps to solve the max-min rate optimization problem \eqref{PDPC}. The initial feasible point $(\btau^{(0)}, \bp^{(0)}, \theta^{(0)}, h^{(0)})$ can be obtained in the same way as described for the NOMA  in Section III-A.
\begin{algorithm}[t]
\begin{algorithmic}[1]
\protect\caption{DPC-based algorithm for max-min rate optimization problem \eqref{PDPC}}
\label{alg2}
\global\long\def\algorithmicrequire{\textbf{Initialization:}}
\REQUIRE  Set $\kappa:=0$ and a feasible point $(\btau^{(0)}, \bp^{(0)}, \theta^{(0)}, h^{(0)})$ for constraints \eqref{cov}, \eqref{P1b}, \eqref{P1c}, and \eqref{P1d}.
\REPEAT
\STATE Solve the convex  optimization problem (\ref{eP1kdpc}) to obtain the optimal solution $(\btau^{(\kappa+1)}, \bp^{(\kappa+1)}, \theta^{(\kappa+1)}, h^{(\kappa+1)})$.
\STATE Set $\kappa:=\kappa+1.$
\UNTIL Convergence\\
\end{algorithmic} \end{algorithm}

\subsection{OMA Algorithm}
The objective function of the OMA-1 problem \eqref{P1oma1} also has similarity in its structure to that for the NOMA problem \eqref{P1}. The non-convex constraint \eqref{cov} can be approximated
by \eqref{ecovk}. Thus, we can use the inequality \eqref{ap2} and the  approximations \eqref{tpk} and \eqref{h1} to approximate the non-concave objective function in \eqref{P1oma1}. Thus, we solve the following convex optimization problem, at the $\kappa$-th iteration, to generate the next iterative feasible point $(\tau^{(\kappa+1)}, p^{(\kappa+1)}, \theta^{(\kappa+1)}, h^{(\kappa+1)})$:
\begin{eqnarray} \label{eP1koma1}
 \ds  \max_{\btau,\mathbf{p},h, \theta} \ f^{\text{OMA-1},(\kappa)}\triangleq \min_{k=1,\dots, K}\ r^{\text{OMA-1},(\kappa)}_k(\btau,\bp,h,\theta)\quad
 \text{s.t.} \quad (\ref{P1b}), (\ref{P1c2}), (\ref{P1d}), (\ref{ecovk}),
\end{eqnarray}
where $r_k^{\text{OMA-1},(\kappa)}(\btau,\bp,h,\theta) = r^{(\kappa)}_{k}(\btau,\bp,h,\theta)$, $\forall$ $k = \{1,\hdots,K\}$ and $r^{(\kappa)}_{k}(\btau,\bp,h,\theta)$ is defined in \eqref{brkk}. Algorithm \ref{alg3} outlines the steps to solve the max-min rate optimization problem \eqref{P1oma1}. The initial feasible point $(\btau^{(0)}, \bp^{(0)}, \theta^{(0)}, h^{(0)})$ can be obtained in the same way as described for the NOMA  in Section III-A.

\begin{algorithm}[t]
\begin{algorithmic}[1]
\protect\caption{OMA-1 algorithm for max-min rate optimization problem \eqref{P1oma1}}
\label{alg3}
\global\long\def\algorithmicrequire{\textbf{Initialization:}}
\REQUIRE  Set $\kappa:=0$ and a feasible point $(\btau^{(0)}, \bp^{(0)}, \theta^{(0)}, h^{(0)})$ for constraints \eqref{cov}, \eqref{P1b}, \eqref{P1c2}, and \eqref{P1d}.
\REPEAT
\STATE Solve the convex  optimization problem (\ref{eP1koma1}) to obtain the optimal solution $(\btau^{(\kappa+1)}, \bp^{(\kappa+1)}, \theta^{(\kappa+1)}, h^{(\kappa+1)})$.
\STATE Set $\kappa:=\kappa+1.$
\UNTIL Convergence\\
\end{algorithmic} \end{algorithm}

Next, in order to solve the OMA-2 problem \eqref{P1oma}, at the $\kappa$-th iteration, we solve the following convex optimization problem to generate the next
iterative feasible point $(\tau^{(\kappa+1)}, p^{(\kappa+1)}, \theta^{(\kappa+1)}, h^{(\kappa+1)})$:
\begin{eqnarray} \label{eP1koma}
 \ds  \max_{\btau,\mathbf{p},h, \theta} \ f^{\text{OMA-2},(\kappa)}\min_{k=1,\dots, K/2}\ r^{\text{OMA-2},(\kappa)}_k(\btau,\bp,h,\theta)\quad
 \text{s.t.} \quad (\ref{P1b}), (\ref{P1c}), (\ref{P1d}), (\ref{ecovk}),
\end{eqnarray}
where
\[
r_k^{\text{OMA-2},(\kappa)}(\btau,\bp,h,\theta) = \begin{cases}
r_{k}^{\text{O},(\kappa)}(\btau,\bp,h,\theta), & \ \  k\in\{1, \dots, K/2\}, \\
r_{j(k)}^{\text{O},(\kappa)}(\btau,\bp,h,\theta), & \ \  j(k)\in\{K/2+1, \dots, K\},
\end{cases}
\]
where $r_{k}^{\text{O},(\kappa)}(\btau,\bp,h,\theta)$ and $r_{j(k)}^{\text{O},(\kappa)}(\btau,\bp,h,\theta)$ are inner approximations (at the $\kappa$-th iteration) of the non-concave functions $r_{k}^{\text{O}}(\btau,\bp,h,\theta)$ and $r_{j(k)}^{\text{O}}(\btau,\bp,h,\theta)$, respectively (defined in \eqref{rkjoma}). Since $r_{j(k)}^{\text{O}}(\btau,\bp,h,\theta)$ is similar to the rate function $r_{j(k)}^{2}(\btau,\bp,h,\theta)$ (defined in \eqref{rk2a} for the NOMA-problem) and $r_{k}^{\text{O}}(\btau,\bp,h,\theta)$ has similar structure too, we can use the inequality \eqref{ap2} (given in the appendix) and the approximations \eqref{pktheta} and \eqref{pik} to find the inner approximations $r_{k}^{\text{O},(\kappa)}(\btau,\bp,h,\theta)$ and $r_{j(k)}^{\text{O},(\kappa)}(\btau,\bp,h,\theta)$.

Thus, by applying the inequality \eqref{ap2} for
\[
\tau=\tau_k, x= \sigma_B \theta /gp_k, y=\tau_k(d_k+h) +g p_{j(k)}/(\sigma_B \theta),
\]
we can obtain the inner approximation for the non-concave rate function $r_{k}^{\text{O}}(\btau,\bp,h,\theta)$, as follows:
\begin{eqnarray}\label{oma1}
r^{\text{O}, (\kappa)}_{k}(\btau,\bp,h,\theta)&\triangleq&\ds
\ta_{k}^{O,(\kappa)}+\tb_{k}^{O,(\kappa)}\left(2-\pikk(\theta,p_{k})-\tilde{\nu}^{O, (\kappa)}_{j(k)}(\tau_k,p_{j(k)},\theta)
\right) -\frac{\tc_{k}^{O,(\kappa)}}{\tau_k},
\end{eqnarray}
where
\begin{equation}\label{oma2}
\tilde{\nu}^{O, (\kappa)}_{j(k)}(\tau_k,p_{j(k)}, \theta)\triangleq
\ds \frac{1}{4}\frac{\left((\tau_k/\tau_k^{(\kappa)})+(d_{k}+h)/(d_{k}+h^{(\kappa)})\right)^2}
{1+gp_{j(k)}^{(\kappa)}/\sigma_B \thetak\tau_k^{(\kappa)}(d_{k}+h^{(\kappa)})}
+\frac{1}{4}\frac{\left((p_{j(k)}/p_{j(k)}^{(\kappa)})+(\thetak/\theta) \right)^2 }{\sigma_B \thetak\tau_k^{(\kappa)}(d_{k}+h^{(\kappa)})/gp_{j(k)}^{(\kappa)}+1},
\end{equation}
and
\begin{eqnarray}
\ta_{k}^{O,(\kappa)}=2\bar{\tau}\ln(1+1/\bar{x}\bar{y}),\
\tb_{k}^{O,(\kappa)}=\frac{\bar{\tau}}{1+\bar{x}\bar{y}},\
\tc_{k}^{O,(\kappa)}=\bar{\tau}^2\ln(1+1/\bar{x}\bar{y}),\label{oma3}
\end{eqnarray}
under
\[
\bar{\tau}=\tau_k^{(\kappa)}, \bar{x}=\sigma_B \thetak/gp_{k}^{(\kappa)}, \bar{y}=\tau_k^{(\kappa)}(d_{k}+h^{(\kappa)})+gp_{j(k)}^{(\kappa)}/(\sigma_B \thetak).
\]
Similarly, by applying the inequality \eqref{ap2} for
\[
\tau=\tau_k, x=\sigma_B \theta/gp_{j(k)}, y=\tau_k(d_{j(k)}+h )+gp_k/(\sigma_B \theta)
\]
we can obtain the inner approximation for the non-concave rate function $r_{j(k)}^{\text{O}}(\btau,\bp,h,\theta)$ as follows:
\begin{eqnarray}\label{oma4}
r^{O, (\kappa)}_{j(k)}(\btau,\bp,h,\theta)&\triangleq&\ds
\ta_{j(k)}^{O,(\kappa)}+\tb_{j(k)}^{O,(\kappa)}\left(2-\pijkk(\theta,p_{j(k)})-\tilde{\nu}^{O, (\kappa)}_{k}(\tau_k,p_{k},\theta)
\right) -\frac{\tc_{j(k)}^{O,(\kappa)}}{\tau_k},
\end{eqnarray}
where
\begin{equation}\label{oma5}
\tilde{\nu}^{O, (\kappa)}_{k}(\tau_k,p_{k}, \theta)\triangleq
\ds \frac{1}{4}\frac{\left((\tau_k/\tau_k^{(\kappa)})+(d_{j(k)}+h)/(d_{j(k)}+h^{(\kappa)})\right)^2}
{1+gp_{k}^{(\kappa)}/\sigma_B \thetak\tau_k^{(\kappa)}(d_{j(k)}+h^{(\kappa)})}
+\frac{1}{4}\frac{\left((p_{k}/p_{k}^{(\kappa)})+(\thetak/\theta) \right)^2 }{\sigma_B \thetak\tau_k^{(\kappa)}(d_{j(k)}+h^{(\kappa)})/gp_{k}^{(\kappa)}+1},
\end{equation}
and
\begin{eqnarray}
\ta_{j(k)}^{O,(\kappa)}=2\bar{\tau}\ln(1+1/\bar{x}\bar{y}),\
\tb_{j(k)}^{O,(\kappa)}=\frac{\bar{\tau}}{1+\bar{x}\bar{y}},\
\tc_{j(k)}^{O,(\kappa)}=\bar{\tau}^2\ln(1+1/\bar{x}\bar{y}),\label{oma6}
\end{eqnarray}
under
\[
\bar{\tau}=\tau_k^{(\kappa)}, \bar{x}=\sigma_B \thetak/gp_{j(k)}^{(\kappa)}, \bar{y}=\tau_k^{(\kappa)}(d_{j(k)}+h^{(\kappa)})+gp_{k}^{(\kappa)}/(\sigma_B \thetak).
\]
Algorithm \ref{alg4} outlines the steps to solve the max-min rate optimization problem \eqref{P1oma}. The initial feasible point $(\btau^{(0)}, \bp^{(0)}, \theta^{(0)}, h^{(0)})$ can be obtained in the same way as described for 
the NOMA  in Section III-A.

\begin{algorithm}[t]
\begin{algorithmic}[1]
\protect\caption{OMA-2 algorithm for max-min rate optimization problem \eqref{P1oma}}
\label{alg4}
\global\long\def\algorithmicrequire{\textbf{Initialization:}}
\REQUIRE  Set $\kappa:=0$ and a feasible point $(\btau^{(0)}, \bp^{(0)}, \theta^{(0)}, h^{(0)})$ for constraints \eqref{cov}, \eqref{P1b}, \eqref{P1c}, and \eqref{P1d}.
\REPEAT
\STATE Solve the convex  optimization problem (\ref{eP1koma}) to obtain the optimal solution $(\btau^{(\kappa+1)}, \bp^{(\kappa+1)}, \theta^{(\kappa+1)}, h^{(\kappa+1)})$.
\STATE Set $\kappa:=\kappa+1.$
\UNTIL Convergence\\
\end{algorithmic} \end{algorithm}

\begin{figure}[t]
    \centering
    \includegraphics[width=0.65 \textwidth]{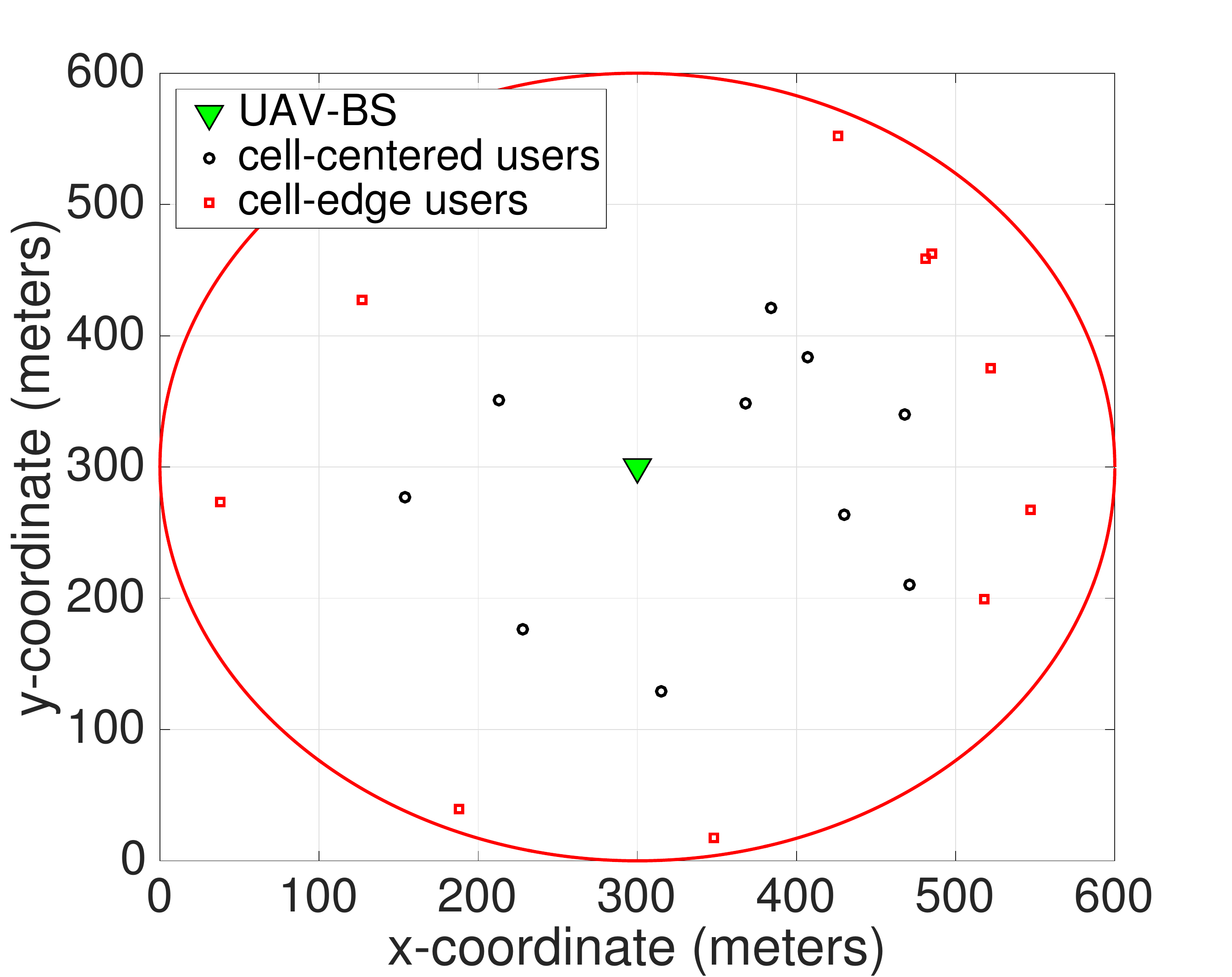} 
  \caption{Network topology used in the simulations.}
    \label{fig:nw_top}
\end{figure}

\section{Simulation Results}

In this section, we analyze the performance of the proposed Algorithms \ref{alg1}-\ref{alg4} via simulations. We use the network topology as used in Fig. \ref{fig:nw_top}, where the cell radius is set to $R = 300$ meters, and there are $K = 20$ users randomly placed within the cell. The UAV BS is at the cell-center and at altitude $\sqrt{h}$ above the ground-level. Fig. \ref{fig:nw_top} shows the ground-level projection of the UAV BS. Half of the users are placed closer to the UAV BS, while the rest of the users are farther from the UAV BS. The channel power gain at a distance of $1$ meter is set to $3.24 \times 10^{-4}$, which incorporates $-38.47$ dB ($1.42 \times 10^{-4}$) path loss and antenna gain $2.2846$ \cite{He-18-Feb-A}. The maximum and minimum UAV altitude are set to $h_\text{max} = 500$ meters and $h_\text{min} = 50$ meters, respectively. The range of the antenna beamwidth is set to $\theta_\text{min} = 0$ and $\theta_\text{max} = \pi/2$ rad. The total power budget is $P = 2$ mW ($3$ dBm). Unless stated otherwise, we set total available bandwidth ${\clB} = 15$ MHz, and the noise power density $\sigma^2 = -174$ dBm/Hz.

\begin{figure}[t]
    \centering
    \includegraphics[width=0.65 \textwidth]{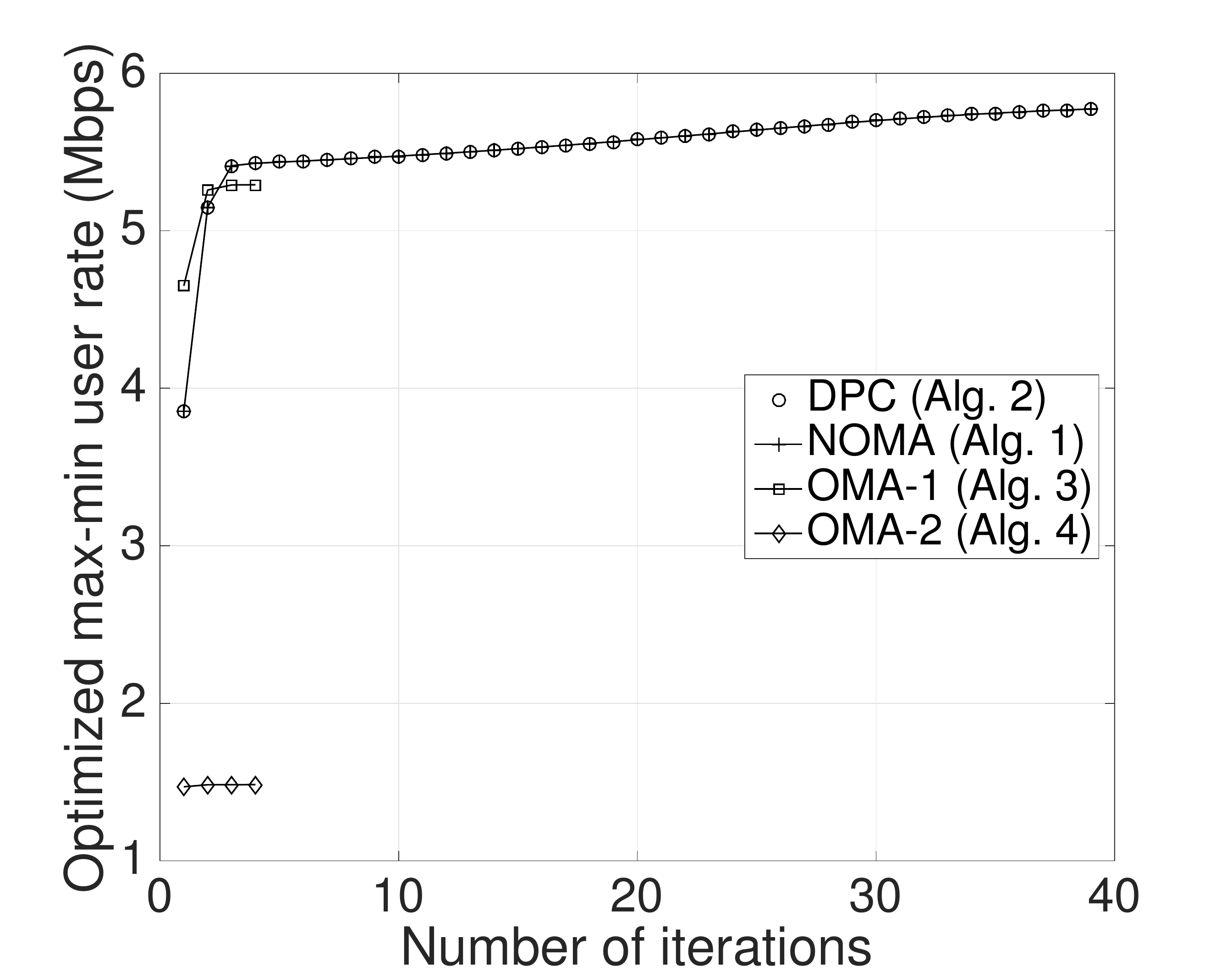} 
  \caption{The convergence of the proposed Algorithms \ref{alg1}-\ref{alg4}.}
    \label{fig:conv}
\end{figure}

\subsection{Performance of the Proposed Algorithms \ref{alg1}-\ref{alg4}}

Fig. \ref{fig:conv} plots the convergence results of the proposed Algorithms \ref{alg1}-\ref{alg4} employing NOMA problem \eqref{P1}, DPC problem \eqref{PDPC}, OMA-1 problem \eqref{P1oma1}, and OMA-2 problem \eqref{P1oma}, respectively.  Fig. \ref{fig:conv} shows that NOMA (Alg. \ref{alg1}) and DPC (Alg. \ref{alg2}) take around $40$ iterations to converge. On the other hand, the convergence of the OMA-1 and OMA-2 (Algorithms \ref{alg3} and \ref{alg4}) requires only four iterations. However, the NOMA and DPC achieve better rates than their OMA counterparts. Even, at the fourth iteration, which is the point where the OMA-1 (Alg. \ref{alg3}) converges, the optimized rate of the NOMA and DPC is better than that of the OMA-1.

\begin{figure*}[t]
    \centering
    \begin{minipage}[h]{0.48\textwidth}
    \centering
    \includegraphics[width=1.01 \textwidth]{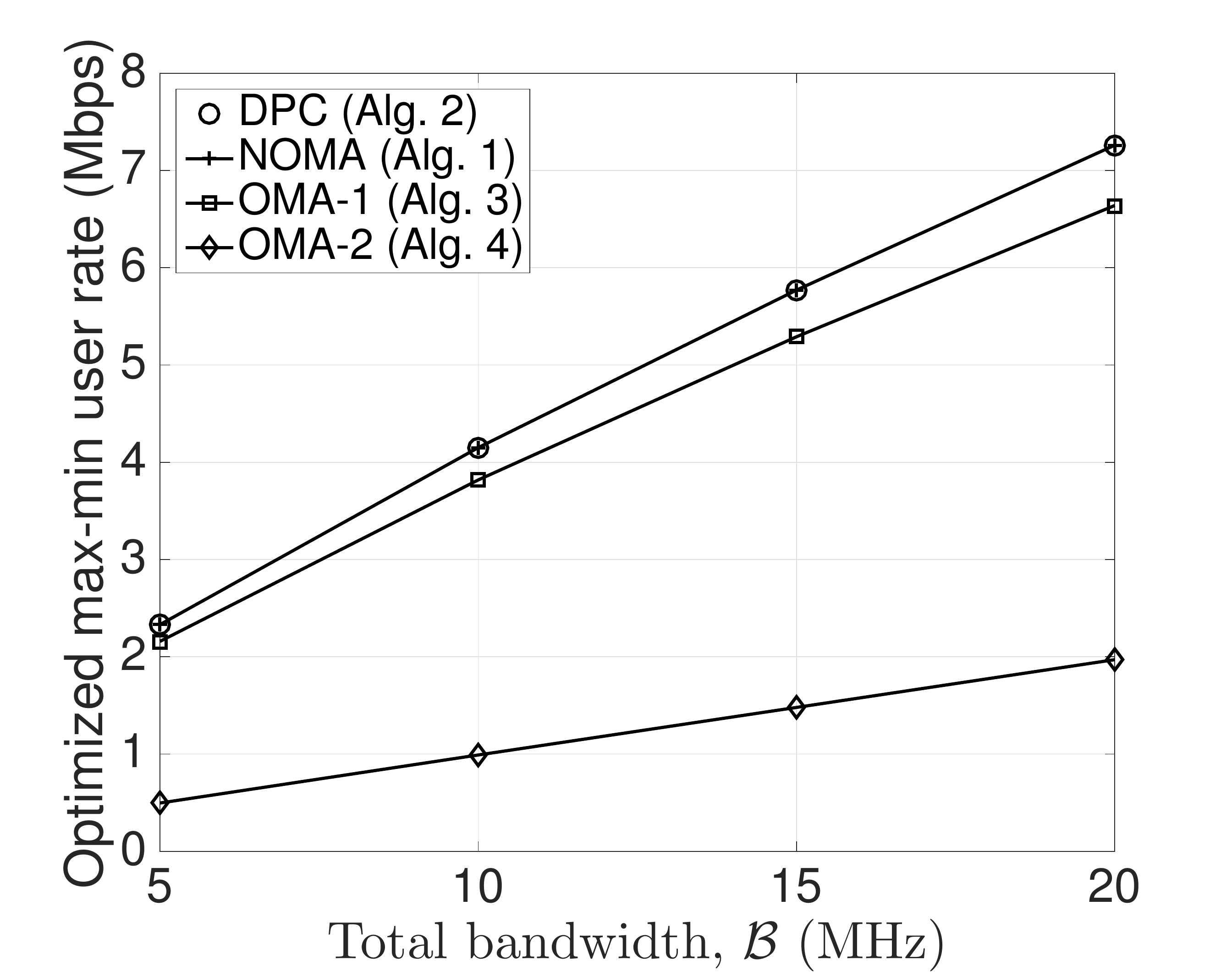}
  \caption{Optimized max-min user rate versus total available bandwidth ${\clB}$, where the noise power density is set to $\sigma^2 = -174$ dBm/Hz.}
  \label{fig:rate_BW}
  \end{minipage}
    \hspace{0.3cm}
    \begin{minipage}[h]{0.48\textwidth}
    \centering
    \includegraphics[width=1.01 \textwidth]{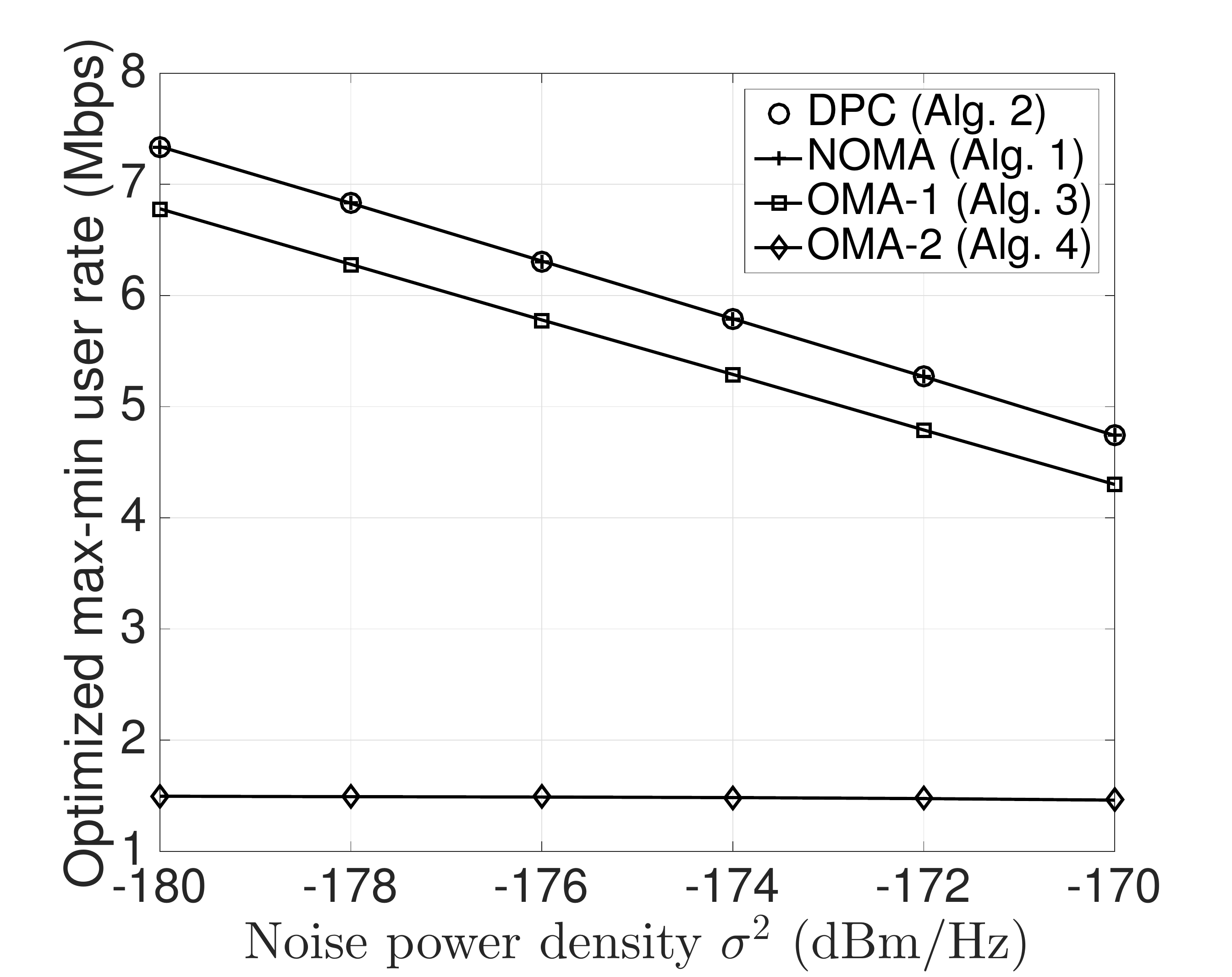}
  \caption{Optimized max-min user rate versus noise power density $\sigma^2$, where the available bandwidth is set to ${\clB} = 15$ MHz.}
  \label{fig:rate_sigma}
  \end{minipage}
\end{figure*}

Fig. \ref{fig:rate_BW} plots the optimized max-min user rate versus the total available bandwidth ${\clB}$. We solve NOMA problem \eqref{P1}, DPC problem \eqref{PDPC}, OMA-1 problem \eqref{P1oma1}, and OMA-2 problem \eqref{P1oma} using Algorithms \ref{alg1}, \ref{alg2}, \ref{alg3}, and \ref{alg4}, respectively. As expected, the optimized rate increases with an increase in the total available bandwidth ${\clB}$. Fig. \ref{fig:rate_BW}  shows that the NOMA and DPC achieve the same performance while clearly outperforming the OMA counterparts. Moreover, we observe that the performance gap between the NOMA and OMA-1 increases with an increase in the available bandwidth $\clB$.

Fig. \ref{fig:rate_sigma} plots the optimized max-min user rates of the proposed Algorithms \ref{alg1}-\ref{alg4} versus the noise power density $\sigma^2$. As expected, the optimized rate decreases with an increase in the noise power density $\sigma^2$.  Fig. \ref{fig:rate_sigma} again shows the same trend that the NOMA and DPC clearly outperform the OMA counterparts. In addition, the performance gap between the NOMA and OMA-1 decreases as the noise power density $\sigma^2$ increases.

\begin{figure*}[t]
    \centering
    \begin{minipage}[h]{0.48\textwidth}
    \centering
    \includegraphics[width=1.01 \textwidth]{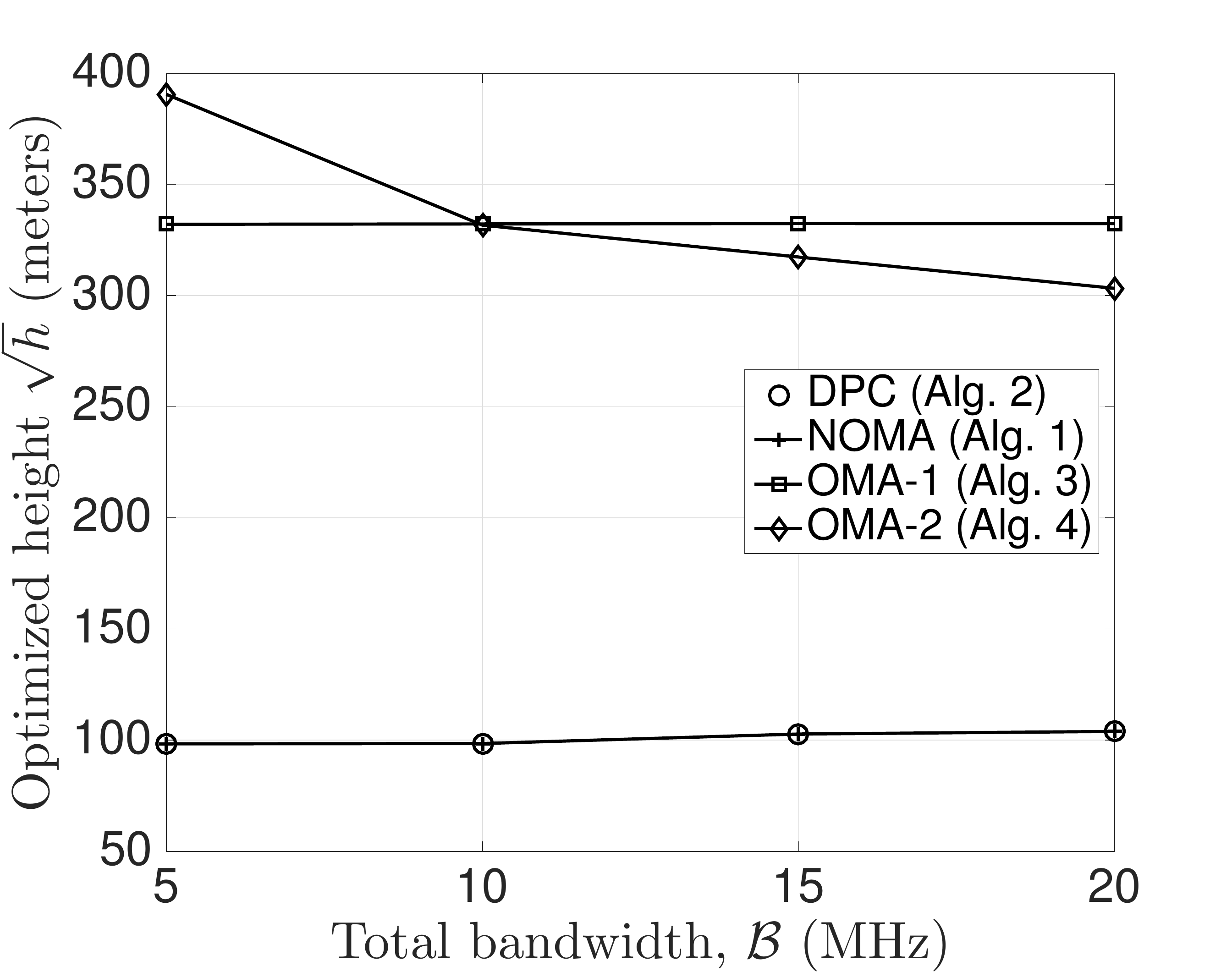}
  \caption{Optimized altitude $\sqrt{h}$ versus total available bandwidth ${\clB}$, where the noise power density is set to $\sigma^2 = -174$ dBm/Hz.}
  \label{fig:height_BW}
  \end{minipage}
    \hspace{0.3cm}
    \begin{minipage}[h]{0.48\textwidth}
    \centering
    \includegraphics[width=1.01 \textwidth]{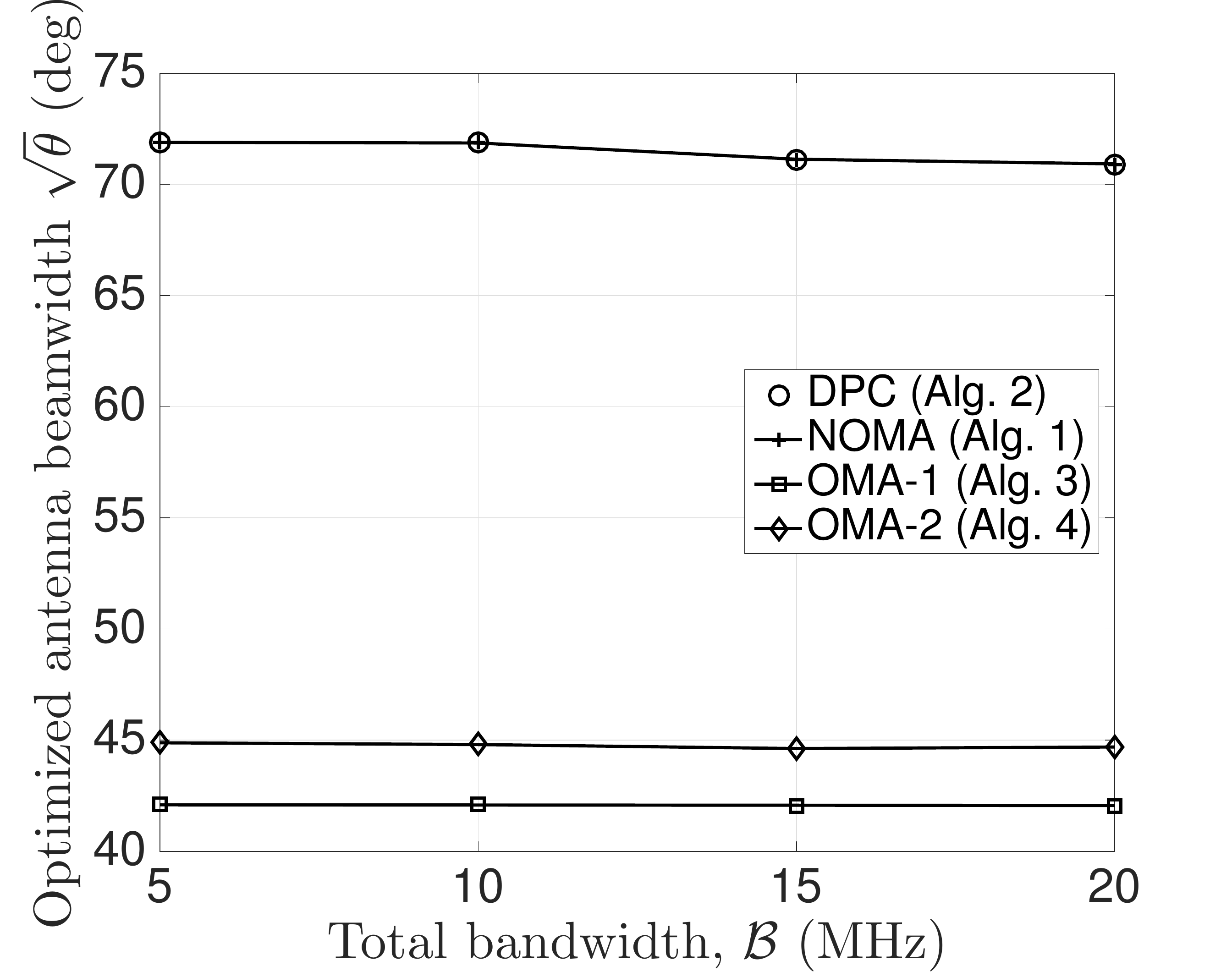}
  \caption{Optimized antenna beamwidth $\sqrt{\theta}$ versus total available bandwidth ${\clB}$, where the noise power density is set to $\sigma^2 = -174$ dBm/Hz.}
  \label{fig:h_BW}
  \end{minipage}
\end{figure*}

Figs. \ref{fig:height_BW} and \ref{fig:h_BW} plot the optimized values of UAV altitude and antenna beamwidth, respectively, after solving all the problems using Algorithms \ref{alg1}, \ref{alg2}, \ref{alg3}, and \ref{alg4}. Figs. \ref{fig:height_BW} and \ref{fig:h_BW} show that there is minor change in the optimized values of the UAV altitude and antenna beamwidth for different values of the total available bandwidth ${\clB}$. This is an interesting and desirable result since the UAV is not required to move much if the bandwidth quota changes.

\begin{figure*}[t]
  \centering
  \subfigure[]
  {
    \hspace*{-0.28in}
    \includegraphics[width=0.35 \textwidth]{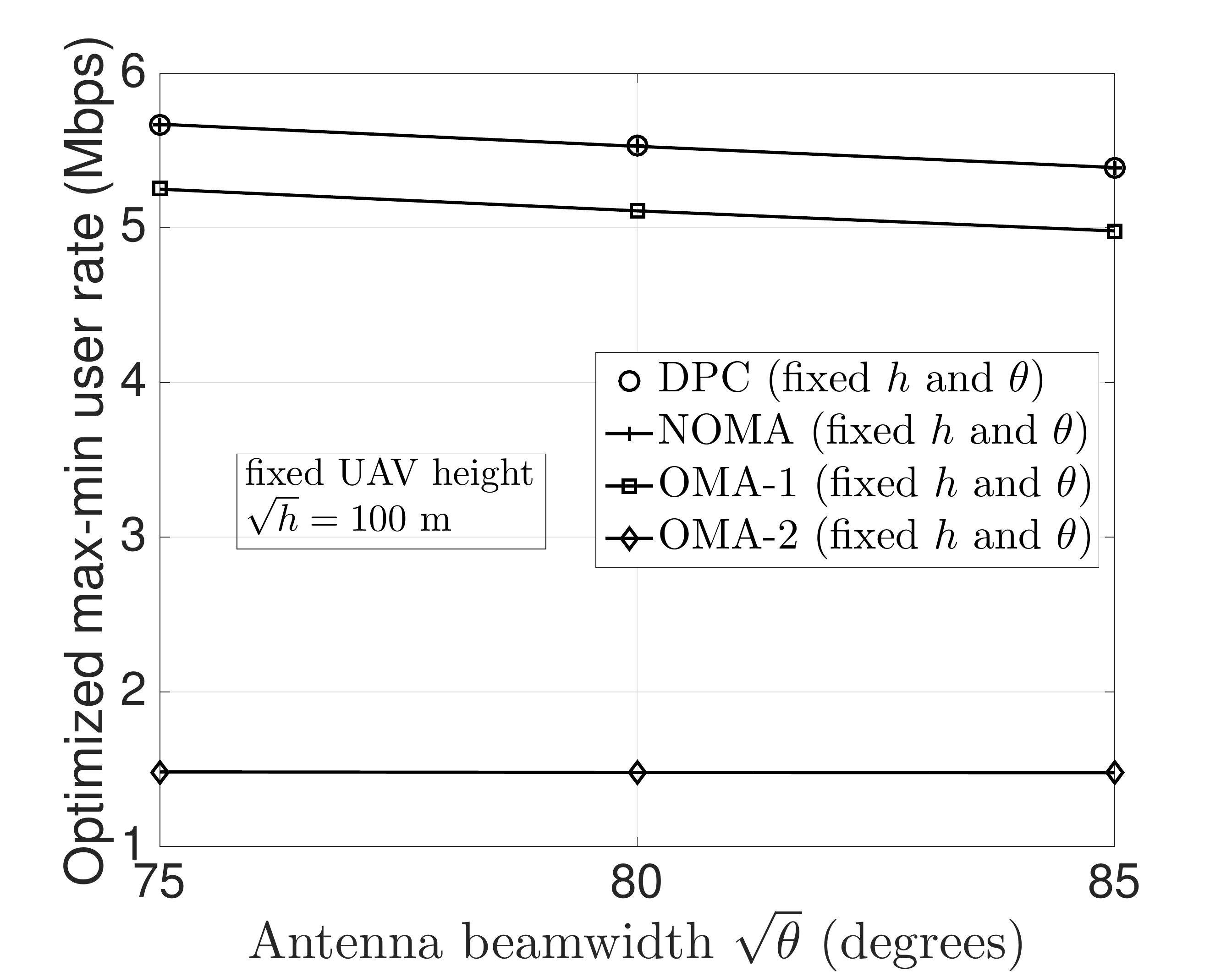}

  }
  \subfigure[]
  {
    \hspace{-0.35in}
    \includegraphics[width=0.35 \textwidth]{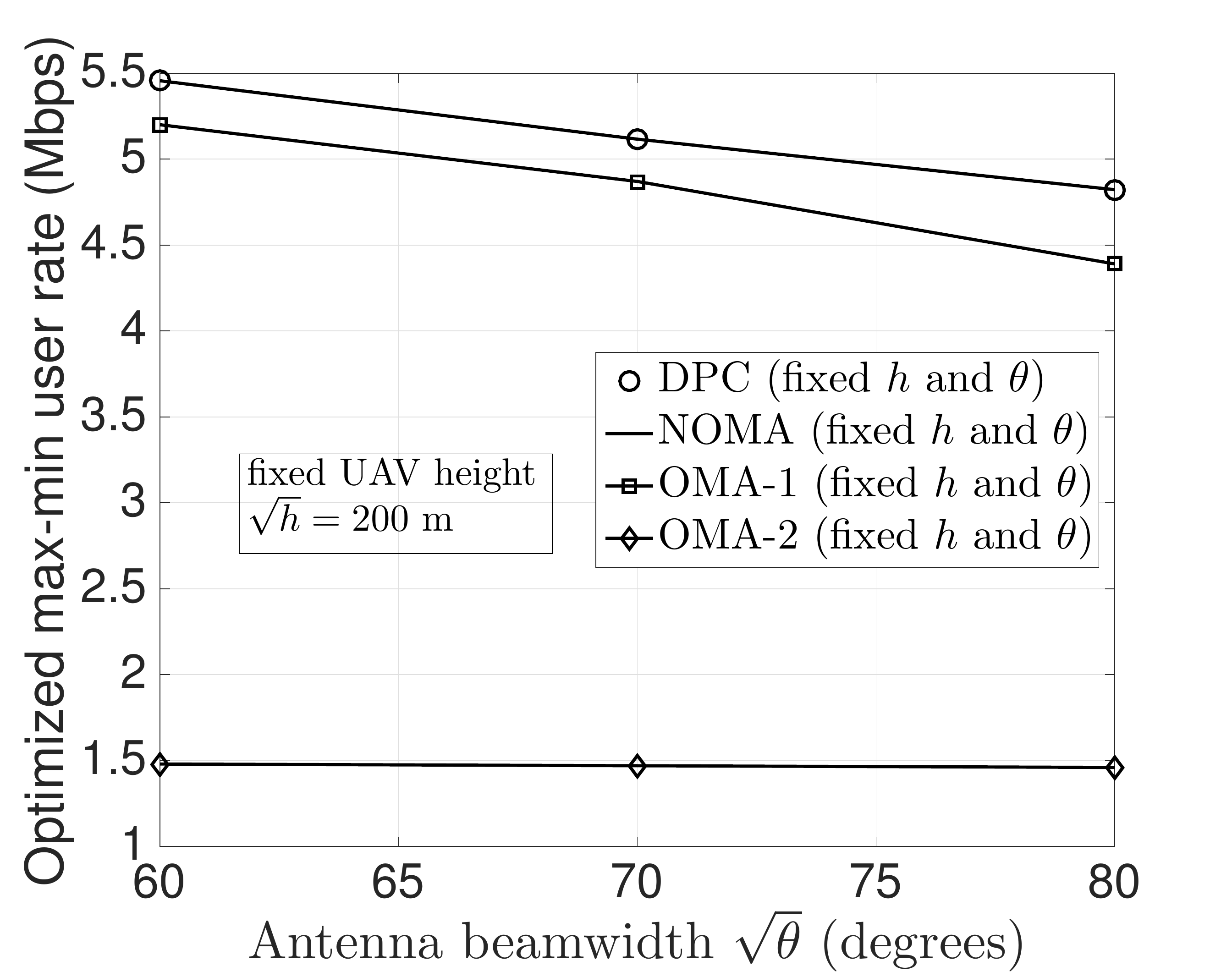}
  }
  \subfigure[]
  {
    \hspace{-0.35in}
    \includegraphics[width=0.35 \textwidth]{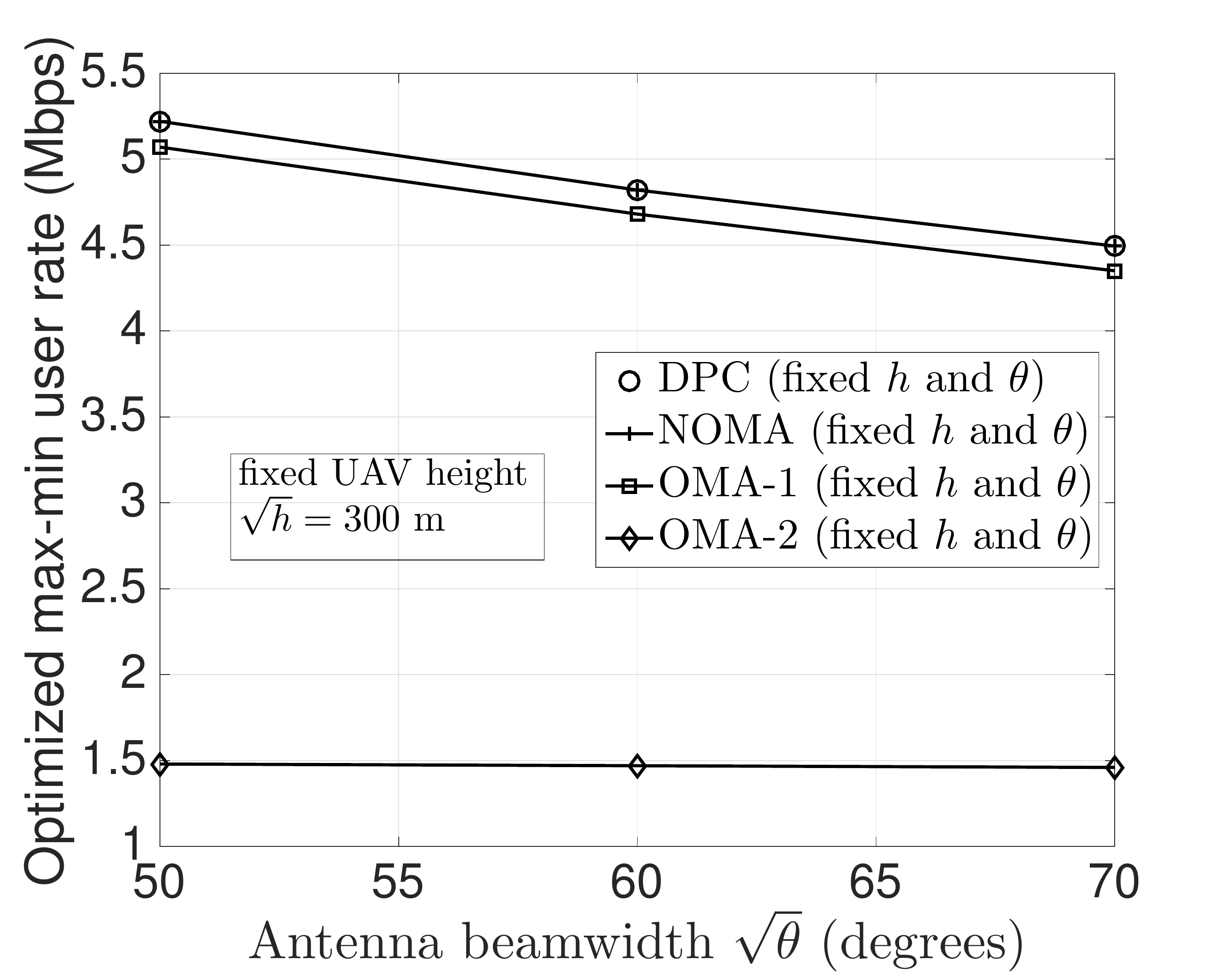}
  }
 \vspace{-0.05in}
  \caption{Optimized max-min user rate under fixed altitude $\sqrt{h}$ and fixed antenna beamwidth $\sqrt{\theta}$, which satisfy \eqref{cov}. Subfig. (a) assumes $\sqrt{h} = 100$ m, subfig. (b) assumes $\sqrt{h} = 200$ m, and subfig. (c) assumes $\sqrt{h} = 300$ m. The total available bandwidth is set to ${\clB} = 15$ MHz. If all parameters including UAV altitude and antenna beamwidth are optimized, as in the proposed algorithm, Fig. \ref{fig:rate_BW} shows that the optimal rate achieved by  NOMA and DPC is $5.77$ Mbps, by OMA-1 is $5.29$ Mbps, and by OMA-2 is $1.48$ Mbps.}
  \vspace{-0.10in}
\label{fig:rate_th}
\end{figure*}

\subsection{Comparison with the Sub-optimal Schemes}

Fig. \ref{fig:rate_th} plots the optimized max-min user rate under fixed altitude $\sqrt{h}$ and fixed antenna beamwidth $\sqrt{\theta}$, such that the constraint \eqref{cov} is satisfied. Again, the bandwidth is set to ${\clB} = 15$ MHz. The Fig. \ref{fig:rate_th}(a) assumes $\sqrt{h} = 100$ m, Fig. \ref{fig:rate_th}(b) assumes $\sqrt{h} = 200$ m, and Fig. \ref{fig:rate_th}(c) assumes $\sqrt{h} = 300$ m. That is, in Fig. \ref{fig:rate_th}, we solve the NOMA problem \eqref{P1}, the DPC problem \eqref{PDPC}, the OMA-1 problem \eqref{P1oma1}, and the OMA-2 problem \eqref{P1oma}, for given fixed altitude $\sqrt{h}$ and fixed antenna beamwidth $\sqrt{\theta}$, i.e., in the absence of constraint \eqref{cov}. Thus, this sub-optimal scheme requires solving only for the optimal power $\bp$ and optimal bandwidth allocation $\btau$. The optimized max-min rates are obviously smaller than the optimized rates as obtained by the proposed optimal Algorithms \ref{alg1}-\ref{alg4} in Fig. \ref{fig:rate_BW}. This is because Algorithms \ref{alg1}-\ref{alg4} jointly optimize all the parameters. In addition, \emph{this justifies the desirability of optimizing UAV-BS altitude and antenna beamwidth.} If all parameters including the UAV altitude and antenna beamwidth are optimized, as in the proposed Algorithms \ref{alg1}-\ref{alg4}, Fig. \ref{fig:rate_BW} shows that the optimal rate achieved by NOMA and DPC is $5.77$ Mbps, by OMA-1 is $5.29$ Mbps, and by OMA-2 is $1.48$ Mbps.

\begin{figure}[t]
    \centering
    \includegraphics[width=0.65 \textwidth]{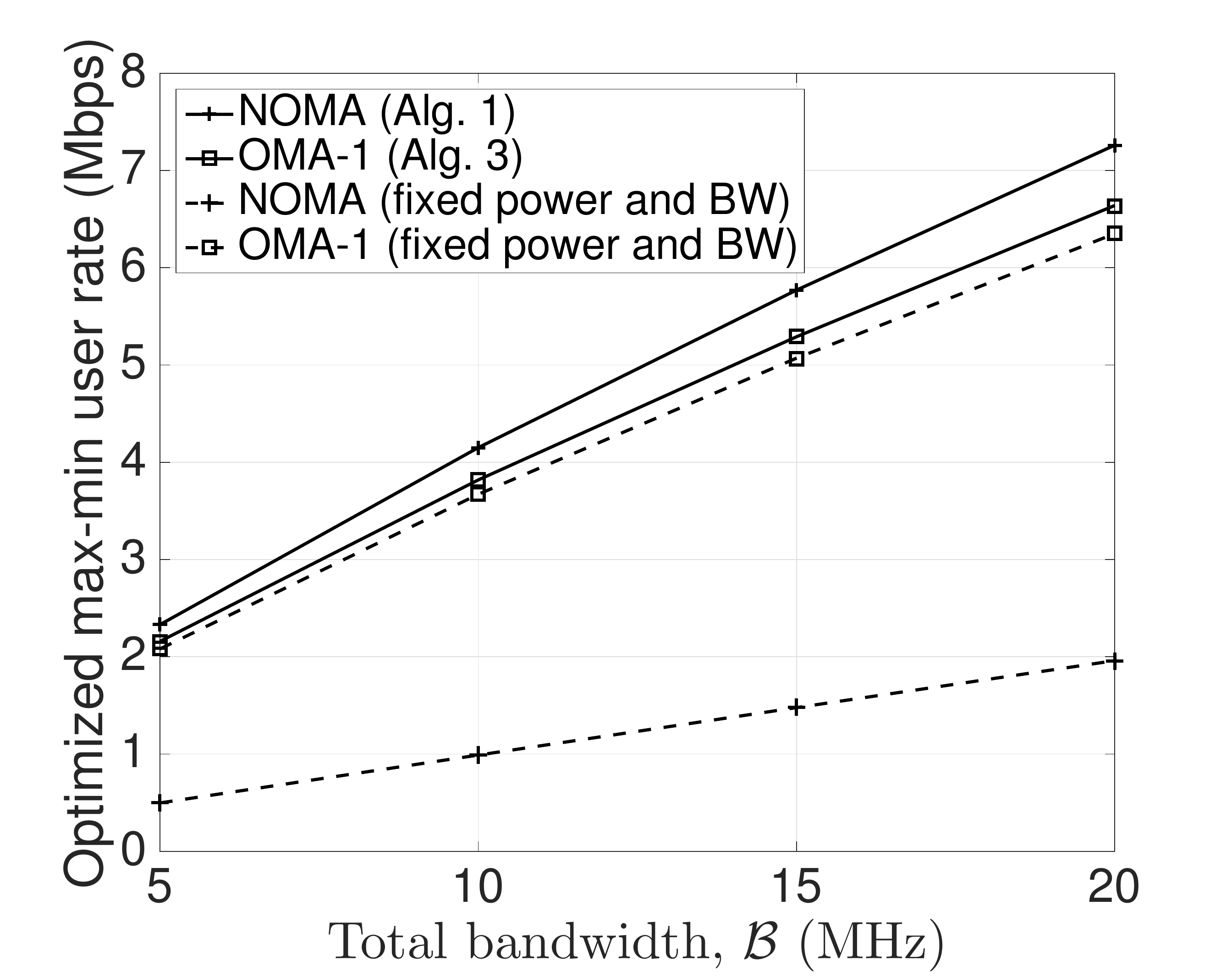} 
  \caption{Comparison of the optimized max-min user rate obtained under fixed power $\bp$ and fixed bandwidth $\btau$ allocation (equal power and equal bandwidth allocation) with the optimized max-min rate achieved by the proposed algorithms.}
    \label{fig:rate_BW_subopt}
\end{figure}

Fig. \ref{fig:rate_BW_subopt} plots the max-min user rate obtained by another sub-optimal scheme, which assumes fixed power $\bp$ and fixed bandwidth $\btau$ allocation and solves  to find the optimal UAV altitude $\sqrt{h}$ and optimal antenna beamwidth $\sqrt{\theta}$. Fig. \ref{fig:rate_BW_subopt} plots results for only NOMA (Alg. \ref{alg1}) and OMA-1 (Alg. \ref{alg3}) because DPC provides similar rate as that obtained by NOMA, and the OMA-2 performs quite poorly. Particularly, we opt for equal power and equal bandwidth allocation, such that, equal power allocaion implies $p_k = P/K$, $\forall$ $k$, while equal bandwidth for NOMA means $\tau_k = 1/(K/2)$, $\forall$ $k \in \{1,\hdots,K/2\}$ and equal bandwidth allocation for OMA-1 means $\tau_k = 1/K$, $\forall$ $k \in \{1,\hdots,K\}$. Fig. \ref{fig:rate_BW_subopt} shows that optimal schemes (Algorithms \ref{alg1} and \ref{alg3} plotted with solid lines) clearly outperform the respective sub-optimal schemes (plotted with dashed lines). Fig. \ref{fig:rate_BW_subopt} shows that sub-optimal NOMA performs quite poorly, and even delivers a worse rate than sub-optimal OMA-1. This is because wise power allocation is necessary for NOMA. On the other hand, the sub-optimal NOMA in Fig. \ref{fig:rate_BW_subopt} assumes equal power allocation, which worsens its achievable rate.

\section{Conclusions}

In this paper, we have considered a UAV-enabled communication network which serves a large number of users by employing NOMA. We have formulated the max-min rate optimization problem under total power, total bandwidth, UAV altitude, and antenna beamwdith constraints. The formulated max-min rate objective function is non-convex in the optimization variables, i.e., the UAV's flying altitude, transmit antenna beamwidth, power allocation and bandwidth allocation for multiple users. We have developed a path-following algorithm to solve the formulated problem. In addition, we have also formulated OMA and DPC-based max-min rate optimization problems and developed respective path-following algorithms to solve them. Finally, our numerical results show that NOMA outperforms OMA and achieves rates similar to those achieved by DPC. Moreover, we have observed a clear rate gain by jointly optimizing all the parameters (power, bandwidth, UAV altitude, and antennas beamwidth), when compared to the case of optimizing subset of these parameters, which confirms the desirability of their joint optimization.

\section*{Appendix: Fundamental Inequalities}
For the convex function $f(x,y,t)\triangleq \ln(1+1/xy)^{1/t}$ \cite{Shetal18}, one has the following
inequality for every $x>0$, $y>0$, $t>0$, $\bar{x}>0$, $\bar{y}>0$ and $\bar{t}>0$:
\begin{eqnarray}
\ds\ds\frac{\ln(1+1/xy)}{t}&=&f(x,y,t)\nonumber\\
&\geq&f(\bar{x},\bar{y},\bar{t})+\la \nabla f(\bar{x},\bar{y},\bar{t}), (x,y,t)-(\bar{x},\bar{y},\bar{t})\ra\nonumber\\
&=&\ds\ds\frac{2\ln(1+1/\bar{x}\bar{y})}{\bar{t}}
+\ds\frac{1}{\bar{t}(1+\bar{x}\bar{y})}(2-x/\bar{x}-y/\bar{y})
-\ds\frac{\ln(1+1/\bar{x}\bar{y})}{\bar{t}^2}t
\label{ap1}
\end{eqnarray}
Therefore, by setting $\tau=1/t$ and $\bar{\tau}=1/\bar{t}$,
\begin{eqnarray}
\ds\tau\ln(1+1/xy)&\geq&\ds 2 \bar{\tau}\ln(1+1/\bar{x}\bar{y})
+\ds\frac{\bar{\tau}}{1+\bar{x}\bar{y}}(2-x/\bar{x}-y/\bar{y})
-\ds\frac{\bar{\tau}^2\ln(1+1/\bar{x}\bar{y})}{\tau}
\label{ap2}
\end{eqnarray}


\end{document}